\documentclass[% 
reprint,
superscriptaddress,
 amsmath,amssymb,amsfonts,
 aps,
 pra,
 longbibliography
]{revtex4-2}

\newcommand{\ket}[1]{\ensuremath{|{#1}\rangle}}
\newcommand{\bra}[1]{\ensuremath{\langle{#1}|}}

\newcommand{\avr}[1]{\ensuremath{\langle{#1}\rangle}}

\newcommand{\cnj}[1]{{#1}^{\ast}}
\newcommand{\hcnj}[1]{{#1}^{\dagger}}
\newcommand{\tcnj}[1]{{#1}^{\mathrm{T}}}

\newcommand{\diag}{\mathop{\rm diag}\nolimits}

\renewcommand{\Re}{\mathop{\rm Re}\nolimits}
\renewcommand{\Im}{\mathop{\rm Im}\nolimits}
\newcommand{\Tr}{\mathop{\rm Tr}\nolimits}

 \newcommand{\vc}[1]{\mathbf{#1}}

 \newcommand{\ind}[1]{\mathrm{#1}}

\newcommand{\prob}{\mathsf{P}}
\newcommand{\dd}{\mathrm{d}}

\usepackage[english]{babel}
\usepackage[utf8x]{inputenc}
\usepackage[T1]{fontenc}

\usepackage{amsmath}
\usepackage{amssymb,amsmath,amsthm,mathtools,mathrsfs}
\usepackage{graphicx}% Include figure files
\usepackage{multirow}% Include figure files
\usepackage{float}% Include figure files
\usepackage{subcaption}% Include figure files
\usepackage{bbm}

\DeclareMathOperator*{\argmax}{arg\,max}

\usepackage{soul,color}
\begin{document}
\DeclareGraphicsExtensions{.png,.pdf}
\title{
  Asymmetry effects in homodyne and heterodyne measurements:
  Positive operator-valued measures and asymptotic security of
  Gaussian-continuous-variable quantum key distribution
}

\author{A. S.~Naumchik}
\email[Email address: ]{naumchik95@gmail.com}
\affiliation{ITMO University, Kronverksky Pr. 49, Saint Petersburg  197101, Russia}

\author{Roman~K.~Goncharov}
\email[Email address: ]{toloroloe@gmail.com}
\affiliation{ITMO University, Kronverksky Pr. 49, Saint Petersburg  197101, Russia}

\author{Alexei~D.~Kiselev}
\email[Email address: ]{alexei.d.kiselev@gmail.com}
\affiliation{%
  Laboratory for Quantum Communications, ITMO University,
  Birzhevaya Line, 16, Saint Petersburg 199034, Russia}
\affiliation{Laboratory of Quantum Processes and Measurements, ITMO
  University, Kadetskaya Line 3b, Saint Petersburg 199034, Russia}

\date{\today}

\begin{abstract}
  We use the Gaussian approximation describing photon count statistics for both the homodyne and the
  double homodyne (heterodyne) measurements to study asymmetry effects arising from imbalance of the
  beam splitters and variations in quantum efficiencies of the photodetectors.  After computing the
  $Q$ symbols of the positive operator-valued measures (POVMs) of noisy measurements that take into
  account the asymmetry effects, the operator representations for the POVMs are obtained in the form
  that assumes applying the additive noise quantum channel to the POVMs of noiseless (ideal)
  measurements.  Additional  noise sources are assumed to be
    represented by statistically independent components included
    into a single excess-noise parameter that enters the CV-QKD analysis.
  For double homodyne detection,
  it was found that the noiseless measurements should generally be expressed in terms of the
  projectors onto squeezed-states and the corresponding squeezed-state operator representation of
  POVM along with the measurement noise channel depend on the squeezing parameter that lies in the
  interval dictated by the condition for the excess noise covariance matrix to be positive
  semidefinite.  The analytical results are used to perform analysis of the asymptotic security of
  the Gaussian-modulated continuous variable quantum key distribution (CV-QKD) protocol in the
  untrusted-noise scenario where the measurement noise is assumed to be accessible to an adversary.
  The inherent non-uniqueness of the operator representation for the double-homodyne POVM manifests
  itself in the squeezing dependent Holevo information whose extrema determine the lower and upper
  bounds for the secret fraction.  For both types of measurements, the mutual information, the
  Holevo information, and the asymptotic secret fraction are sensitive to asymmetry effects, leading
  to degraded performance of the protocol.
\end{abstract}
 
 \maketitle

%%%%%%%%%%%%%%
\section{Introduction}
\label{sec:intro}
%%%%%%%%%%%%%%

Optical homodyne detection is a
well-established experimental technique of quantum optics for
measuring phase-sensitive properties of
light fields.
In a typical four-port homodyne setup
described in textbooks~(see, e.g., the textbook~\cite{Vogel:bk:2006}),
a weak quantum signal is superimposed with a strong local oscillator (LO) at a beam splitter
and the resulting output fields are detected
with photodiodes,
so that information
about the properties of the signal field quadratures
is extracted from the statistics of the difference photocurrent.

In the realm of continuous-variable (CV) quantum systems,
this technique plays a key role as a fundamentally important measurement method 
and,
over the course of the last two decades of the past century,
homodyne-based measurement schemes
have been the subject of numerous
theoretical
studies~\cite{Yuen:ieee:1978,Yuen:ieee:1979,Yuen:ieee:1980,Yuen:ol:1983,
  Walker:jmo:1987,Collett:jmo:1987,Braunstein:pra:1990,Vogel:pra:1993,Leonhardt:pra:1993,Ban:jmo:1997,Banaszek:pra:1997,Kuzmich:jmo:1998,Feng:josab:2016}
reviewed in~\cite{Welsch:progopt:1999}.
In the limit of strong local oscillator,
the noiseless homodyne detection becomes a projective measurement
described by projectors onto a suitable set of quadrature eigenstates,
whereas a similar result for the eight-port double homodyne scheme,
which is known to be equivalent to
the so-called heterodyne detection scheme,
is represented by the coherent states.
In this case,
homodyne and double homodyne (heterodyne)
schemes can be regarded as implementations
of Gaussian measurements defined as measurement procedures that
produce a Gaussian probability distribution of outcomes for any
Gaussian state~\cite{Giedke:pra:2002,Fiurasek:pra:2007}
(see also the monograph~\cite{Serafini:bk:2023}).
Note that the
LO intensity and fluctuations
induced effects emerging beyond the strong LO approximation
are studied in
Refs.~\cite{Vogel:pra:1995,Cives:joptb:2000,Auyuanet:josab:2019,Olivares:pla:2020,Thekkadath:pra:2020}. 

A wide variety of technologically important applications,
where homodyne related techniques
are of crucial importance,
include the quantum tomography of
CV states~\cite{Leonhardt:bk:1997,Smithey:prl:1993,Kuhn:pra:1995,DAuria:prl:2009,Lvovsky:rmp:2009,Buono:josab:2010,Teo:pra:2017,Tiunov:optica:2020,Olivares:njp:2019,Bing:optexp:2020,Kumar:pra:2022,Oh:npjqi:2019}
and
quantum communication protocols~\cite{Braunstein:rmp:2005,Semenov:pra:2012}
such as the CV quantum teleportation~\cite{Furusawa:sci:1998,Shi:lpor.2023}
and continuous-variable quantum key
distribution (CV-QKD)
systems~\cite{Grosshans:prl:2002,Weedbrook:prl:2004,Weed:rmp:2012,Zhang:apr:2024}.

Practical security
for real-world QKD systems
requires that
deviations between the system and the
ideal be assessed 
to take into account
the imperfections of the devices~\cite{Scarani:rmp:2009,Xu:rmp:2020}.
Imperfections of the photodetectors
such as
non-unity detector efficiencies,
electronic noise, dark counts, finite bandwidth and
photon-number resolution, detector dead times
and afterpulses
introduce excess
noise that affects CV-QKD
security~\cite{Chi:njp:2011,Jouguet:pra:2012,Wang:prappl:2019,Lupo:prxq:2022,Semenov:pra:2024,hajomer2025finite,Oruganti:qst:2025}.
Noisy homodyne and heterodyne measurements are
no longer projective and have to be analyzed in terms of
the positive operators-valued measures (POVMs).
Imperfections leading to measurement noise
may open side channels exploitable by an adversary
using wavelength-dependent or detector-blinding
attacks~\cite{huang2012wavelength,huang2014quantum,qin2018homodyne,qin2016quantum}.
Countermeasures to such attacks include spectral filtering, detector balancing, and
careful calibration of the photodetectors~\cite{Zou:optexp:2022,Wang:jlt:2023,Pan:pra:2025}.

In this paper, we focus our attention
on the quantum measurement imperfections
that will be collectively referred to as the \textit{asymmetry effects}.
These effects arise from the unbalanced beam splitters
and the difference in quantum efficiencies of the photodetectors.
Although in recent studies~\cite{Barchielli:pra:2022,Ruiz:heliyon:2023,Bartlett:25}, asymmetry effects
are found to be of importance for quantum random number
generators~\cite{Barchielli:pra:2022}
and security analysis of CV-QKD~\cite{Ruiz:heliyon:2023,Bartlett:25},
a systematic POVM-based theoretical analysis has not yet received proper attention.
Our goal is to fill the gap.
To this end, we derive the POVMs and
study how the asymmetry effects influence
the asymptotic security
of the Gaussian CV-QKD protocol
in the untrusted-noise
scenario where an adversary controls measurement
imperfections.

The paper is organized as follows.
In Sec.~\ref{sec:homodyne}, we
obtain the statistical distribution of difference photon counts
with the asymmetry effects taken into account
and apply the Gaussian approximation to deduce the $Q$ symbol of
the corresponding homodyne POVM.
The general strategy described in Sec.~\ref{subsec:hom-POVM}
is used to deduce
the operator representation for the POVM. 
In Sec.~\ref{sec:double-homodyne},
the eight-port double-homodyne scheme
is analyzed using the method presented in the previous section.
After deriving the $Q$ symbol of POVM in the Gaussian approximation,
we show that the operator representation
of the POVM can be obtained by
applying the
additive noise channel with the squeezing dependent
excess noise covariance matrix
to the noiseless squeezed-state measurements.
Thus, it turned out that the operator
representation of the double-homodyne POVM
is non-unique as it depends on the squeezing parameter.
In
Sec.~\ref{sec:protocol}, we apply our results for homodyne and
double-homodyne detection to evaluate the asymptotic secret fraction
for the Gaussian-modulated coherent-state (GMCS) CV-QKD protocol
in the untrusted-noise scenario.
It is found that, for heterodyne detection,
the Holevo information requires additional optimization
with respect to the squeezing parameter.
Finally, in Sec.~\ref{sec:conclusion}, we summarize the main results and outline the directions for
future research.
Technical details are relegated to the Appendices:
Appendix~\ref{sec:appendix_numerical} details analysis of
the accuracy of the Gaussian approximation against the exact Skellam distributions;
Appendix~\ref{sec:appendix_comparison}
examines alternative approximations using Bessel-function asymptotics of the Skellam
distribution;
Appendix~\ref{sec:MI-and-EBtoPm} derives analytical expressions for mutual information
under asymmetric detection and
proves the equivalence between the preparation and measurement-based
and entanglement-based protocol representations.

\begin{figure}
    \centering
    \includegraphics[width=0.6\linewidth]{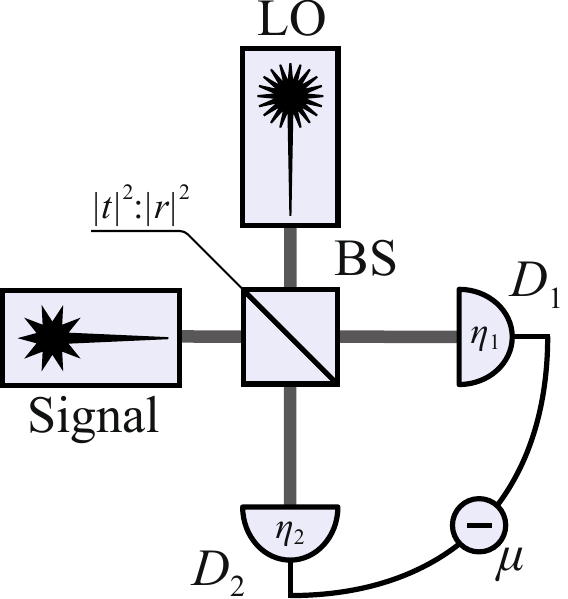}
    \caption{Scheme of a homodyne receiver: S is the source of the
      signal mode with the annihilation operator $\hat{a}$,
      LO is the source of the reference mode (local oscillator) with the annihilation
      operator $\hat{a}_{L}$, and BS is the beam splitter with
      the amplitude transmission and reflection coefficients
      $t=\cos\theta$ and $r=\sin\theta$, respectively; photodetectors $D_1$ and $D_2$ have quantum efficiencies
     $\eta_{1}$ and $\eta_{2}$, and $\mu\equiv m_1-m_2$ is the photon count
      difference.}
    \label{fig:homodyne}
\end{figure}

%%%%%%%%%%%%%%%%%%%%%%%%%%%%
\section{Homodyne detection}
\label{sec:homodyne} 
%%%%%%%%%%%%%%%%%%%%%%%%%%%

\begin{figure*}
    \centering
    \begin{subfigure}[]{.45\textwidth}
\includegraphics[width=\linewidth]{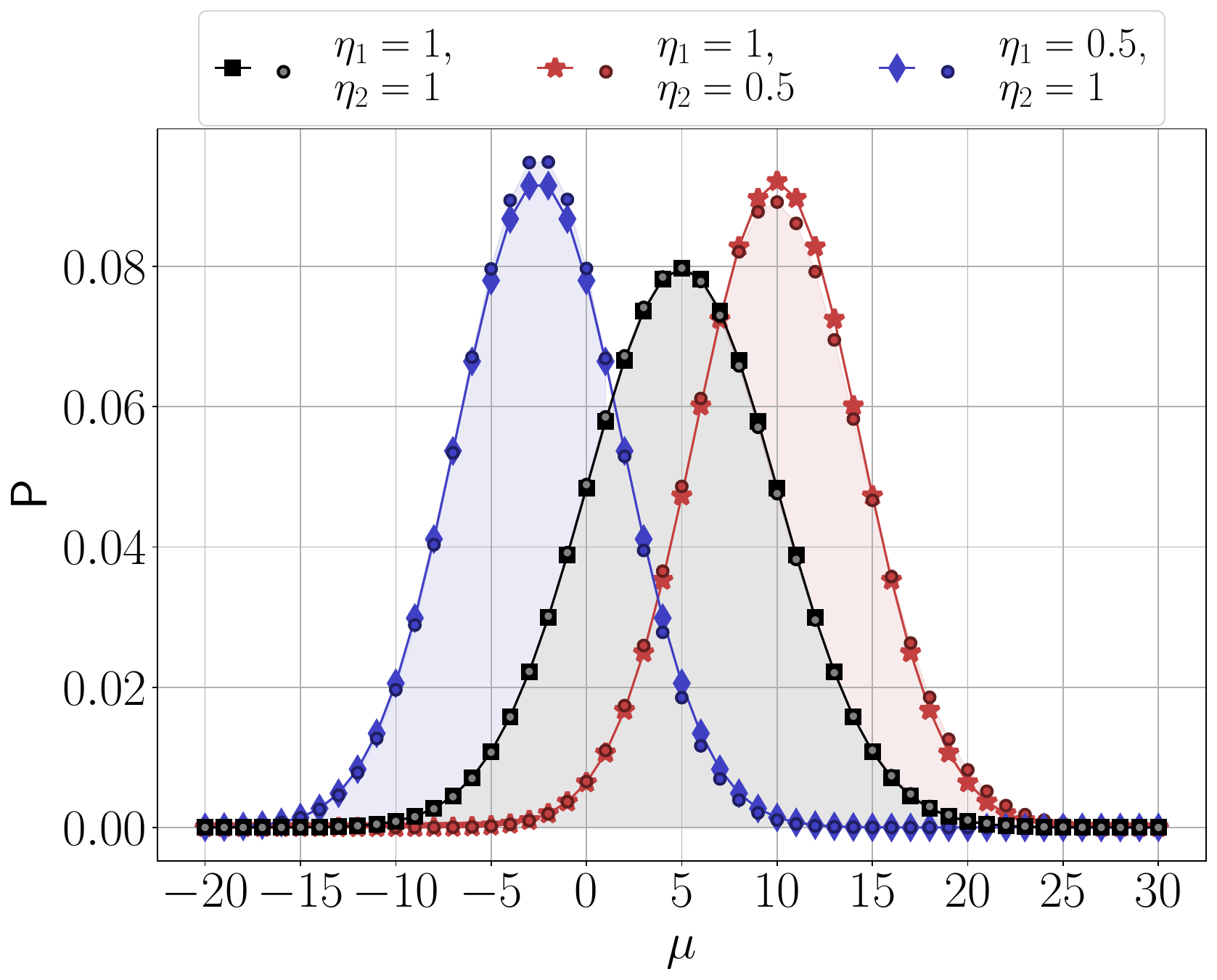}
\caption[]{$\ket{\psi}=\ket{\alpha},\:$$\alpha=0.5$}
\label{fig:dist_alp}
        \end{subfigure}
        \hfill
        \begin{subfigure}[]{.45\textwidth}
 \includegraphics[width=\linewidth]{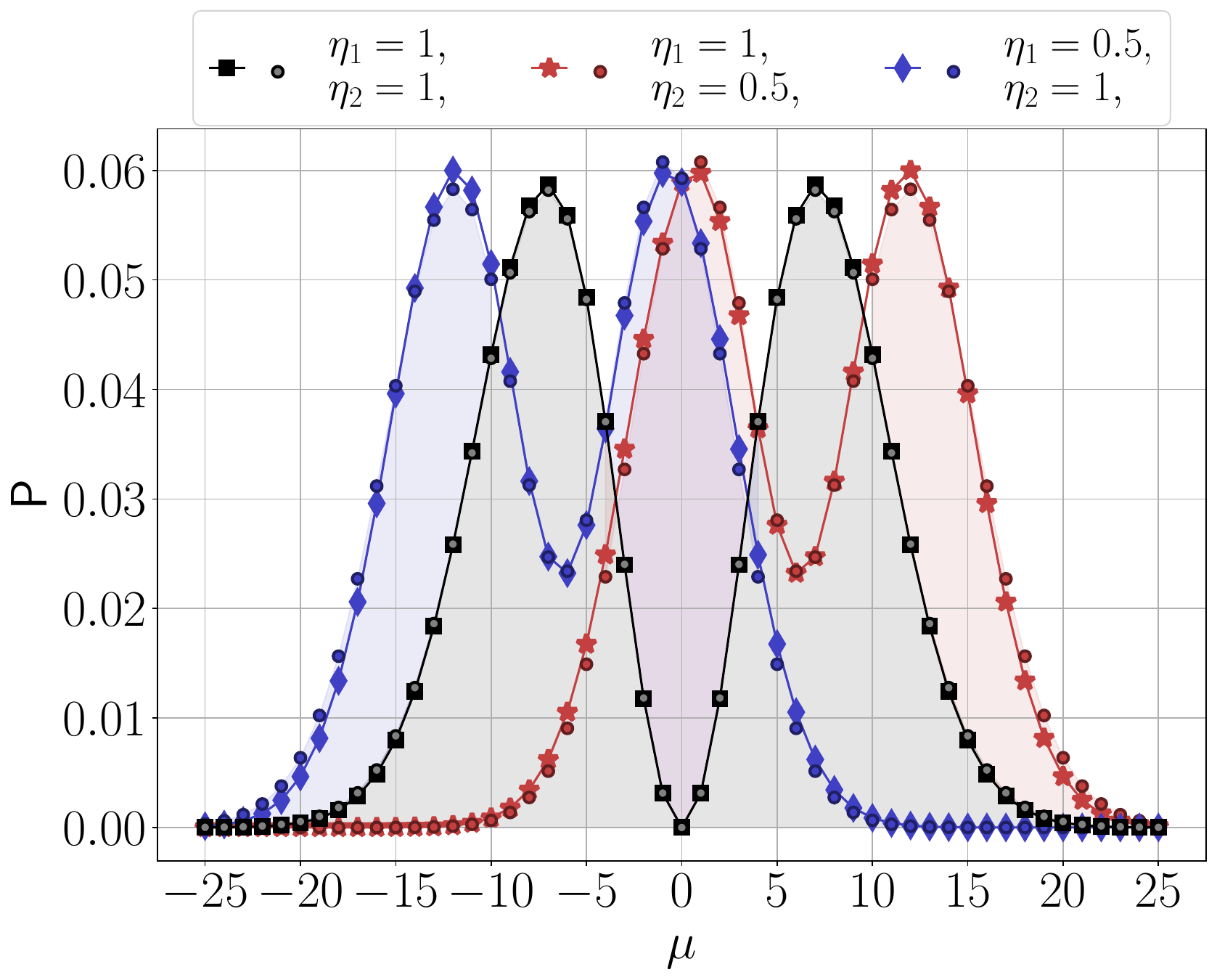}
\caption[]{$\ket{\psi}=\ket{n},\:n=1$}
\label{fig:dist_fock}
\end{subfigure}
\caption{Exact (circle dots) and approximate (solid lines with
  markers) statistical distributions of photon count difference for
  the signal mode prepared in (a)~the coherent state and in (b)~the
  single photon Fock state computed for for different efficiencies at
  $\left|\alpha_L\right|=5$ and balanced beamspliiter.  }
\label{fig:dist-H}
\end{figure*}

  In this section, we remind the reader
  about the basic steps of the
  theoretical treatment of the homodyne measurements
and deduce the operator representation of the POVM
  needed for subsequent analysis of GMCS CV-QKD
  in the untrusted noise scenario.
  Our anaysis is confined to the effects of imbalance of beam splitter and mismatch in quantum efficiencies
  of the photodetectors that, as is expected, will produce asymmetry-induced shift and excess noise
  in the statistical distribution of difference photon counts.
  Following the approach widely used in the security analysis of
  CV-QKDs~(see, e.g., Ref.~\cite{Laudenbach:aqt:2018}),
  we assume that other sources of imperfections can be taken into account
  through the excess-noise parameter introduced in Sec.~\ref{sec:protocol}. 

%%%%%%%%%%%%%%%%%%%%%%%%%%%%
\subsection{Gaussian approximation}
\label{subsec:hom-gauss}
%%%%%%%%%%%%%%%%%%%%%%%%%%%%

We begin with a brief discussion
of the homodyne measurement setup
schematically depicted in Fig.~\ref{fig:homodyne}.
To this end, we assume that the beam spitter is unbalanced
and its scattering matrix is chosen to be a real-valued
rotation matrix with the transmission and reflection amplitudes,
$t$ and $r$, given by
\begin{align}
  \label{eq:BS-amplitudes}
  t=\cos\theta\equiv C,
  \quad
  r=\sin\theta\equiv S.
\end{align}
Then the input coherent states
of the signal mode and the local oscillator 
are transformed into the output coherent states
as follows
\begin{align}
  \label{eq:BS}
  &
    \ket{\alpha,\alpha_L}\mapsto\ket{\alpha_1,\alpha_2},
  \\
  &
    \label{eq:amplitudes}
    \alpha_1=C\alpha+S\alpha_L,\quad
    \alpha_2=-S\alpha+C\alpha_L,
\end{align}
so that
the joint probability of $m_1$ and $m_2$ photon counts
for the photodetectors $D_1$ and $D_2$
can be computed from the well-known Kelley-Kleiner formula~\cite{Kelley:pr:1964}
(see also Ref.~\cite{Vogel:pra:1993}):
\begin{align}
    \label{eq:1}
    \prob(m_1,m_2)&=
    \bra{\alpha_1,\alpha_2}:\prod_{l=1}^{2}\frac{(\eta_l\hat{n}_l)^{m_l}e^{-\eta_l\hat{n}_l}}{m_l!}:\ket{\alpha_1,\alpha_2}
    \notag
  \\
  &
 =\prod_{l=1}^{2}\frac{(\eta_l\left|\alpha_L\right|^2)^{m_l}}{m_l!}e^{-\eta_l\left|\alpha_L\right|^2}    
\end{align}
where
$:\ldots:$ stands for normal ordering,
index $l\in\{1,2\}$ labels the output ports of the beam splitter,
$\hat{n}_l=\hcnj{\hat{a}}_l\hat{a}_l$ is the photon number operator,
$m_l$ is the number of photon counts,
$\eta_l$ is the quantum efficiency of the detector $D_l$.
We can now introduce the photon count difference
\begin{equation}
  \label{eq:dleta-m}
    \mu = m_1-m_2
  \end{equation}
  so that its statistical distribution %of the photon count difference
  can be written in
  the form of a product of the two Poisson distributions as follows 
\begin{align}
\label{eq:poisson}  
    \prob(\mu) &=\sum_{m_2=\max(0,-\mu)}^{\infty}
    \frac{(\eta_1|\alpha_1|^2)^{\mu+m_2}}{(\mu+m_2)!}e^{-\eta_1|\alpha_1|^2}
    \notag
  \\
  &
    \times
\frac{(\eta_2|\alpha_2|^2)^{m_2}}{m_2!}e^{-\eta_2|\alpha_2|^2}.
\end{align}
It is well known that, by performing a summation on $m_2$,
the probability $\prob(\mu)$ reduces to
the Skellam distribution given by~\cite{skellam1946frequency}
\begin{align}
  \label{eq:accurate}
  \prob(\mu)&=e^{-\eta_1|\alpha_1|^2}e^{-\eta_2|\alpha_2|^2}
    \left(\frac{\eta_1|\alpha_1|^2}{\eta_2|\alpha_2|^2}\right)^{\mu/2}
    \notag
  \\
  &
    \times
I_{\mu}\left(2\sqrt{\eta_1\eta_2|\alpha_1|^2|\alpha_2|^2}\right),
\end{align}
where $I_k(z)$ is the modified Bessel function of the first kind~\cite{NIST:hndbk:2010}.

An important point is that,
at sufficiently large
$|\alpha_1|$ and $|\alpha_2|$,
Poisson distributions that enter Eq.~\eqref{eq:1}
can be approximated using
the probability density functions of normal distributions
with mean and variance equal to the mean of
the corresponding Poisson distribution, $\lambda_i=\eta_i |\alpha_i|^2$.
Then, in the continuum limit where
the summation in Eq.~\eqref{eq:poisson} is replaced by integration,
the Skellam distribution~\eqref{eq:accurate}
can be approximated assuming
that the amplitude of the local oscillator, $\left|\alpha_L\right|$, is large
(the strong-LO approximation) and
we can apply the convolution formula
for Gaussian probability densities
\begin{align}
  &
  \label{eq:convolution}
  \int G(x_1-x_2;\sigma_1)G(x_2;\sigma_2)\dd
    x_2=G(x_1;\sigma_1+\sigma_2),
  \\
  &
    \label{eq:G-notation}
    G(x;\sigma)\equiv\frac{1}{\sqrt{2\pi\sigma}}\exp\Bigl(-\frac{x^2}{2\sigma}\Bigr).
\end{align}

The above procedure
immediately leads to the Gaussian approximation
of the following form:
\begin{align}
  &
  \label{eq:Gaussian}
    \prob_G(\mu)=G(\mu-\mu_G;\sigma_G),
    \\
  &
    \label{eq:sigma_G}
    \sigma_G=\eta_1|\alpha_1|^2+\eta_2|\alpha_2|^2
    \approx
    (\eta_1S^2+\eta_2C^2) \left|\alpha_L\right|^2,
  \\
  &
    \label{eq:mu_G}
    \mu_G=\eta_1|\alpha_1|^2-\eta_2|\alpha_2|^2
    \approx
    (\eta_1S^2-\eta_2C^2) \left|\alpha_L\right|^2
    \notag
  \\
  &
    +CS (\eta_1+\eta_2) \left|\alpha_L\right| \avr{\hat{x}_\phi}
\end{align}
where
\begin{align}
  &
  \label{eq:avr-x-phi}
  \avr{\hat{x}_\phi}\equiv \bra{\alpha}\hat{x}_\phi\ket{\alpha}={2\Re\alpha e^{-i\phi}},
  \quad \phi=\arg\alpha_L
\end{align}
is the average of
the phase-rotated quadrature operator of the signal mode
given by
\begin{equation}
\hat{x}_\phi=\hat{a}e^{-i\phi}+\hat{a}^\dag e^{i\phi}.
    \label{eq:quad-op}
\end{equation}

Alternatively, the probability~\eqref{eq:Gaussian}
can be rewritten in the form
\begin{align}
  \label{eq:Pgood-w-def}
{\prob}_G(x)\equiv {\prob}_G^{(\ind{H})}(x)=\frac{1}{\sqrt{2\pi\sigma_G}}
    \exp \left[-\frac{(x-\avr{\hat{x}_\phi})^2}{2\sigma_x}\right],
\end{align}
where
$x$ is the quadrature variable given by
\begin{align}
  \label{eq:x-def}
  x\equiv\frac{\mu}{(\eta_1+\eta_2)CS\left|\alpha_L\right|}-\frac{\eta_1S^2-\eta_2C^2}{\left(\eta_1+\eta_2\right)CS}\left|\alpha_L\right|
\end{align}
and $\sigma_x$ is the quadrature variance
\begin{align}
  &
        \label{eq:sigma-x}
    \sigma_x\equiv
    \frac{\eta_1S^2+\eta_2C^2}{
    \left[(\eta_1+\eta_2)CS\right]^2
    }.
\end{align}
It is rather straightforward to
minimize the variance~\eqref{eq:sigma-x} with respect to the transmittance $C^2$. The result is
\begin{align}
  \label{eq:sigma-x-min}
  &\min_{C^2}\sigma_{x}=\left(
\frac{\sqrt{\eta_1}+\sqrt{\eta_2}}{\eta_1+\eta_2}
  \right)^2\ge 1.%,\\
  %\label{eq:C-min}
  %&\argmin_{C^2}\sigma_x=\frac{\sqrt{\eta_1}}{\sqrt{\eta_1}+\sqrt{\eta_2}}.
\end{align}

%%%%%%%%%%%%%%%%%%%%%%%%%%%%
\subsection{Positive operator-valued measure}
\label{subsec:hom-POVM}
%%%%%%%%%%%%%%%%%%%%%%%%%%%%

Our next step is to construct
the positive operator-valued measure
(POVM) based on the Gaussian approximation $\prob_G$.
To this end, we note that the probability~\eqref{eq:Pgood-w-def}
is the expectation value of the POVM in the coherent state
given by
\begin{align}
  \label{eq:PG-as-avr}
    \prob_G=\bra{\alpha}\hat{\Pi}_G\ket{\alpha}.
\end{align}

The operator $\hat{\Pi}_G$ can be regarded as a non-normalized density
matrix describing a Gaussian state~\cite{Serafini:bk:2023}.
In the single mode case,
such state is generally
characterized by the $2\times 2$ covariance matrix
$\Sigma_m$ and the vector of averaged quadratures
$\vc{r}_m$.
So, the $Q$ symbol of $\Pi_G$
is determined by
the Gaussian shaped distribution of the form:
\begin{align}
  \label{eq:PG-Q-symbol}
    \bra{\alpha}\hat{\Pi}_G\ket{\alpha}
    \propto
    &
    \exp\{-2 (\vc{r}-\vc{r}_m)^{\ind{T}}
    \notag
    \\
&
  \times  (\Sigma_m+\mathbbm{1})^{-1}(\vc{r}-\vc{r}_m)\},
  \\
    &
      \alpha=\alpha_1+i \alpha_2,
      \quad \vc{r}=
      \begin{pmatrix}
        \alpha_1\\ \alpha_2
      \end{pmatrix},
\end{align}
where
$\mathbbm{1}$ is the identity matrix
and the superscript $\ind{T}$ stands for transposition,
so that the transpose of $\vc{r}$ is the row vector:
$\tcnj{\vc{r}}=(\alpha_1,\alpha_2)$.

It is known that POVMs of noisy measurements can be introduced
using the dual of a quantum channel acting on
POVMs representing perfect (noiseless)
measurements~\cite{Serafini:bk:2023}.
In what follows, we shall use
the classical mixing (additive noise) quantum channel
\begin{align}
  \label{eq:Phi_N}
  &
    \Phi_N:
    \hat{\Pi}_{\ind{id}}\mapsto
    \hat{\Pi}_G= \Phi_N(\hat{\Pi}_{\ind{id}})=
    \frac{2}{\pi\sqrt{\det\Sigma_N}}
    \notag
  \\
  &
    \times \int_{\mathbb{C}}\dd^2\beta
    \exp\{-2 \vc{s}^T\Sigma_N^{-1}\vc{s}\}
    \hat{D}(\beta)\hat{\Pi}_{\ind{id}}\hat{D}(-\beta),
\end{align}
where
$\tcnj{\vc{s}}=(\Re(\beta), \Im(\beta))=(\beta_1,\beta_2)$,
$\dd^2\beta=\dd \beta_1\dd \beta_2$
and $\hat{D}(\beta)$ is the displacement operator given by
\begin{align}
    \label{eq:D-def}
\hat{D}(\beta) =
e^{\beta\hcnj{\hat{a}}-\beta^*\hat{a}}.
\end{align}
This is a self-dual channel,
$\cnj{\Phi}_N=\Phi_N$,
that transforms the POVM of the ideal projective Gaussian measurements,
$\hat{\Pi}_{\ind{id}}$,
into the POVM of the noisy measurements,
$\hat{\Pi}_{G}$, 
by changing the covariance matrix $\Sigma_{\ind{id}}$
as follows
\begin{align}
  \label{eq:Sigma_N}
  \Phi_N:
  \Sigma_{\ind{id}}\mapsto
  \Sigma_G=\Sigma_{\ind{id}}+\Sigma_N,
  \quad \Sigma_N\ge 0
\end{align}
where $\Sigma_N$ is
the excess noise covariance matrix.

Our general strategy involves the following steps:
(a)~using the $Q$ symbol of POVM to derive the covariance matrix $\Sigma_m$;
(b)~identifying the states representing ideal measurements
and $\Sigma_{\ind{id}}$;
and (c)~using the additive noise channel with the excess noise covariance matrix $\Sigma_N=\Sigma_m-\Sigma_{\ind{id}}$ to construct the POVM of noisy measurements.

In the case of
the perfectly symmetric noiseless homodyne measurement with
$\eta_1=\eta_2=1$ and $C=S=1/\sqrt{2}$,
the average~\eqref{eq:PG-as-avr}
takes the form
\begin{equation}
    \label{eq:P_0}
    \prob_{\ind{id}}^{(\ind{H})}=
    \frac{1}{\left|\alpha_L\right|} Q_{x,\phi}(\alpha), 
  \end{equation}
  where
  $x=\mu/\left|\alpha_L\right|$
  and $Q_{x,\phi}(\alpha)$ is the Husimi $Q$ distribution
for the eigenstate of the phase-rotated quadrature
operator~\eqref{eq:quad-op},
$\ket{x,\phi}$,
given by (see, e.g., the textbook~\cite{Vogel:bk:2006})
\begin{align}
  &
    \label{eq:Q-x}
  Q_{x,\phi}(\alpha)=
    |\langle \alpha|x , \phi \rangle |^2 =G(x-\avr{\hat{x}}_\phi;1).
\end{align}
Thus, we are led to the well-known result that, in the Gaussian approximation,
the POVM describing sharp homodyne measurements
\begin{align}
  &
  \label{eq:POVM-P0}
    \hat{\Pi}_{\ind{id}}^{(\ind{H})}=\frac{1}{\left|\alpha_L\right|}|x,\phi\rangle\langle x,\phi|
\end{align}
 is proportional to
the projector onto $\ket{x,\phi}$.

Since the eigenstate of the rotated quadrature
$\hat{x}_\phi$ can be represented as
the squeezed coherent state
taken in the limit of infinitely large squeezing 
\begin{align}
  &
  \label{eq:ket-x-phi}
  \ket{x,\phi}=\hat{R}(\phi)\ket{x},
  \quad
  \hat{R}(\phi)=\exp\{i\phi\hcnj{\hat{a}}\hat{a}\},
  \\
  &
    \label{eq:ket-x}
  \ket{x}=\lim_{r\to+\infty}\sqrt{\frac{\cosh(r)}{(2\pi)^{1/2}}}\hat{D}(x/2)\hat{S}(-r)\ket{0},
\end{align}
where $\hat{S}(r)$ is the squeezing operator given by
\begin{align}
\label{eq:S-def}
    \hat{S}(\zeta)=e^{\frac{1}{2}\left(\zeta\hat{a}^{\dag2}-\zeta^*\hat{a}^2\right)},
\end{align}
its covariance matrix assumes a similar form~\cite{Serafini:bk:2023}:
\begin{align}
  \label{eq:Sigma-H-0}
  \Sigma_{\ind{id}}^{\ind{(H)}}=\lim_{r\to+\infty}R(\phi)\,\ind{diag}(e^{-2r},e^{2r})\,\tcnj{R}(\phi),
\end{align}
where $R_\phi$ is the rotation matrix given by
\begin{align}
  \label{eq:rotation-matrix}
  R(\phi)=\begin{pmatrix}
    \cos\phi&-\sin\phi\\
    \sin\phi&\cos\phi
  \end{pmatrix}.
\end{align}

In a more general asymmetric case with $\eta_1\ne \eta_2$
and $C\ne S$, from the Q symbol~\eqref{eq:Pgood-w-def}
we deduce the following expression for the covariance matrix of
noisy homodyne measurements
\begin{align}
  &
    \label{eq:Sigma-H-m}
    \Sigma_{m}^{(\ind{H})}= \lim_{r\to+\infty}R(\phi)\,\ind{diag}(e^{-2r}+\sigma_N,e^{2r})\,\tcnj{R}(\phi)
    \notag
  \\
  &
=\Sigma_{\ind{id}}^{(\ind{H})}+\Sigma_{N}^{(\ind{H})},
  \\
  &
    \label{eq:Sigma-N-H}
\Sigma_{N}^{(\ind{H})}=R(\phi)\,\ind{diag}(\sigma_N,0)\,\tcnj{R}(\phi),
\end{align}
where $\sigma_N$ is the excess noise given by
\begin{align}
  &
  \label{eq:sgm_N}
     \sigma_N=\sigma_x-1\ge 0.
\end{align}
Note that the excess noise variance $\sigma_N$ takes into account asymmetry effects
and its non-negativity is guarantied by inequality~\eqref{eq:sigma-x-min}.

We can now use formula~\eqref{eq:Phi_N}
along with the algebraic identities
\begin{align}
  \label{eq:phi_N-rel1}
  &
    \int_{\mathbb{C}}\dd^2\beta
    \exp\{-2 \vc{s}^T\bigl[R(\phi)\Sigma_N^{-1}\tcnj{R}(\phi)\bigr]
    \vc{s}\}
    \hat{D}(\beta)\hat{\Pi}_{\ind{id}}\hat{D}(-\beta)
    \notag
  \\
  &
    =\int_{\mathbb{C}}\dd^2\beta
    \exp\{-2 \vc{s}^T\Sigma_N^{-1}\vc{s}\}
    \hat{D}(\beta e^{i\phi})\hat{\Pi}_{\ind{id}}\hat{D}(-\beta e^{i\phi}),
  \\
  &
    \label{eq:phi_N-rel2}
    \hat{D}(\beta e^{i\phi})\hat{R}(\phi)\hat{D}(\alpha_m)=
    \hat{R}(\phi)\hat{D}(\beta)\hat{D}(\alpha_m)
\end{align}
to derive the POVM in the form of randomly displaced
projectors
\begin{align}
\label{eq:homodyne-povm}    
  \hat{\Pi}_G^{(\ind{H})}
  &
               =\frac{1}{(\eta_1+\eta_2)CS\left|\alpha_L\right|}
    \notag
  \\
  &
    \times
    \int \dd x' G\left(x-x'; \sigma_N\right) \ket{x',\phi}\bra{x',\phi}.
\end{align}
Alternatively, this result can be obtained
using the convolution identity~\eqref{eq:convolution}
that leads to the relation
\begin{align}
  &
  \label{eq:theform}
  \prob_G^{(\ind{H})}(x)=\sqrt{\frac{\sigma_x}{\sigma_G}} \int \mathop{\dd x'} G(x-x';\sigma_N) Q_{x',\phi}(\alpha)  
\end{align}
linking the Q symbols of
the projectors $\ket{x',\phi}\bra{x',\phi}$
and the POVM $\hat{\Pi}_G^{(\ind{H})}$.

%   This result immediately gives
%   a general formula for the Gaussian approximation POVM 

% In the limiting case of perfect homodyne,
% we have
% \begin{align}
%   \label{eq:lim}
%     \lim_{\sigma_x \to 1}G(x;\sigma_N)=\lim_{\sigma_N \to 0}G(x;\sigma_N)=\delta(x),
% \end{align}
% where $\delta(x)$ is the Dirac $\delta$-function,
% which is the expected behavior for
% Eq.~\eqref{eq:theform} to hold.
% Therefore, the constructed POVM~\eqref{eq:homodyne-povm}
% is well-defined for all possible parameters of the homodyne scheme.

% From Eq.~\eqref{eq:Pgood-w-def} % and Eq.~\eqref{eq:Sigma-H-0} 
% we are also able to construct POVM's~\eqref{eq:homodyne-povm} covariance matrix:
% \begin{align}
%   \label{eq:Pi_h}
%   \Sigma^{\ind{H}}&=\lim_{z\to0}R_\phi^{\ind{T}}\left[\ind{diag}(z,z^{-1})\right]R_\phi+\sigma_N\mathbbm{1}
%                     \notag
%   \\
%                   &
%                     \equiv\Sigma_0^{\ind{H}}+\Sigma_N^{\ind{H}}, 
% \end{align}
% where $\Sigma_N^{\ind{H}}$ denotes the covariance matrix of the associated excess noise:
% \begin{align}
%     \label{eq:def-sigma_h-N}
%   &\Sigma^{\ind{H}}_N=\begin{pmatrix}
%         \sigma_N&0\\
%         0&\sigma_N
%     \end{pmatrix}.
% \end{align}
% % \textit{The ability of the homodyne scheme to reconstruct the quadrature of the state is well known
% % (see e.g. Ref.~\cite{liu2021homodyne}), which is showcased by Eq.~\eqref{eq:homodyne-povm}.
% % }

The exact and approximate analytical results for photon count difference statistical distributions, given by Eq.~\eqref{eq:accurate} and Eq.~\eqref{eq:Gaussian} respectively, are valid
for the case where the LO and signal modes are both
in the coherent states. In the more general case where the quantum state of the signal mode
is $\ket{\psi}$,
the probability distributions can be evaluated
using the relations
\begin{align}
  &
\label{eq:average-P-func}
    \prob(\mu;\ket{\psi})=\int  P_{\ket{\psi}}(\alpha)\prob(\mu;\alpha)\dd^2\alpha,
    \notag
  \\
  &
    \prob_G(x;\ket{\psi})=\bra{\psi}\hat{\Pi}_G\ket{\psi},
\end{align}
where $P_{\ket{\psi}}(\alpha)$ is the Glauber $P$ function of the quantum state
$\ket{\psi}$.
In Fig.~\ref{fig:dist-H},
we show the results computed for the single-photon Fock state
obtained utilizing
the well-known expression for
the $P$-function of Fock states $\ket{\psi}=\ket{n}$ given by
%\cite{vogel2006quantum}
\begin{equation}
\label{eq:P-function_fock}
P_{\ket{n}}(\alpha)=
\frac{e^{|\alpha|^2}}{n!}
\Bigl(
\frac{\partial^2}{\partial\alpha\partial\cnj{\alpha}}
\Bigr)^n
\delta^{2}(\alpha),
\end{equation}
where $\delta^{2}(\alpha)=\delta(\Re\alpha)\delta(\Im\alpha)$.

Figure~\ref{fig:dist-H} displays
the photocount difference probabilities
computed from
the exact and Gaussian probability distributions
for the balanced beam splitter at different photodetector efficiencies
and signal mode input states.
Fig.~\ref{fig:dist_alp} shows that,
in agreement with Eq.~\eqref{eq:mu_G},
the asymmetry in photodetection
results in a shift of the probability maximum.
%Note that, at $\delta\theta=0$, the photocount variance~\eqref{eq:sigma_G}, $\sigma_G=(\eta_1+\eta_2)\left|\alpha_L\right|^2/2$, and the quadrature variance~\eqref{eq:sigma-x}, $\sigma_x=2/(\eta_1+\eta_2)$, are both invariant under transposition of the photodetectors: $\eta_{1}\leftrightarrow \eta_{2}$.

The distributions for the single-photon states
are depicted in Fig.~\ref{fig:dist_fock}
and, similar to the coherent state,
demonstrate the effect of asymmetry-induced shift.
Another noticeable effect is that
the probability minimum between the central and side peaks
become less pronounced.

From Fig.~\ref{fig:dist-H} it becomes apparent that the performance of the Gaussian approximation worsens in
the presence of asymmetry, which we will quantify in Appendix~\ref{sec:appendix_numerical}.

Our concluding remark concerns an alternative method
to approximate Eq.~\eqref{eq:accurate}
with a Gaussian-shaped distribution
that is based on the asymptotic expansions of the modified Bessel functions.
In Appendix~\ref{sec:appendix_comparison} we show that, for the asymmetric homodyne scheme,
this method generally leads to ill-posed POVMs because
the corresponding quadrature variance
appears to be too small leading to a negative
contribution of excess noise.

\begin{figure}
    \centering
    \includegraphics[width=.85\linewidth]{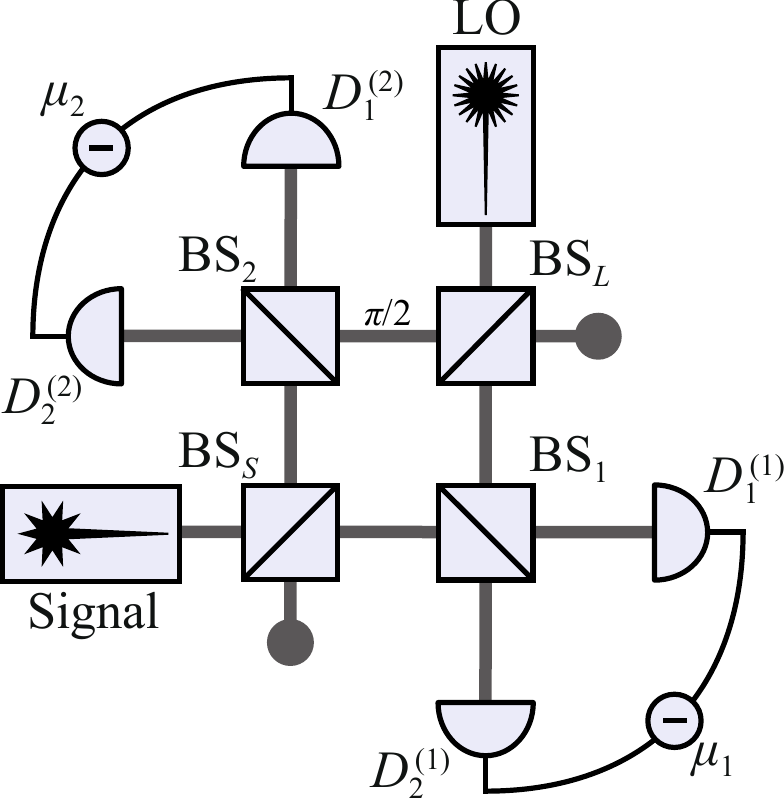}
    \caption{Scheme of an eight port double-homodyne receiver, which consists of two homodyne setups, each composed of a beamsplitter, two photodetectors, and a photocurrent subtractor.
      S is the source of the signal mode
      $\hat{a}$; LO is the source of the reference mode $\hat{a}_{L}$;
      BS$_S$ (BS$_L$) is the signal mode (local oscillator) beam
      splitter
      with transmission and reflection coefficients $t_S$ ($t_L$) and $r_S$ ($r_L$), respectively;
      $\frac{\pi}{2}$ is the quarter wave phase shifter;
BS$_i$ is the beam splitter of $i$th homodyne with transmission and reflection coefficients $t_i$ and $r_i$;
$D_{1,2}^{(i)}$ are the photodetectors of the $i$th homodyne with quantum detection efficiency $\eta_{1,2}^{(i)}$;
and $\mu_i=m_1^{(i)}-m_2^{(i)}$ is the photon count difference registered by
the detectors of $i$th homodyne, $i\in\{1,2\}$.
    }
    \label{fig:double-homodyne}
\end{figure}

%%%%%%%%%%%%%%%%%%%%%%%%%%%%
\section{Double homodyne detection}
\label{sec:double-homodyne}
%%%%%%%%%%%%%%%%%%%%%%%%%%%%%%

\begin{figure*}
    \centering
    \begin{subfigure}[c]{.3\linewidth}
\includegraphics[width=\linewidth]{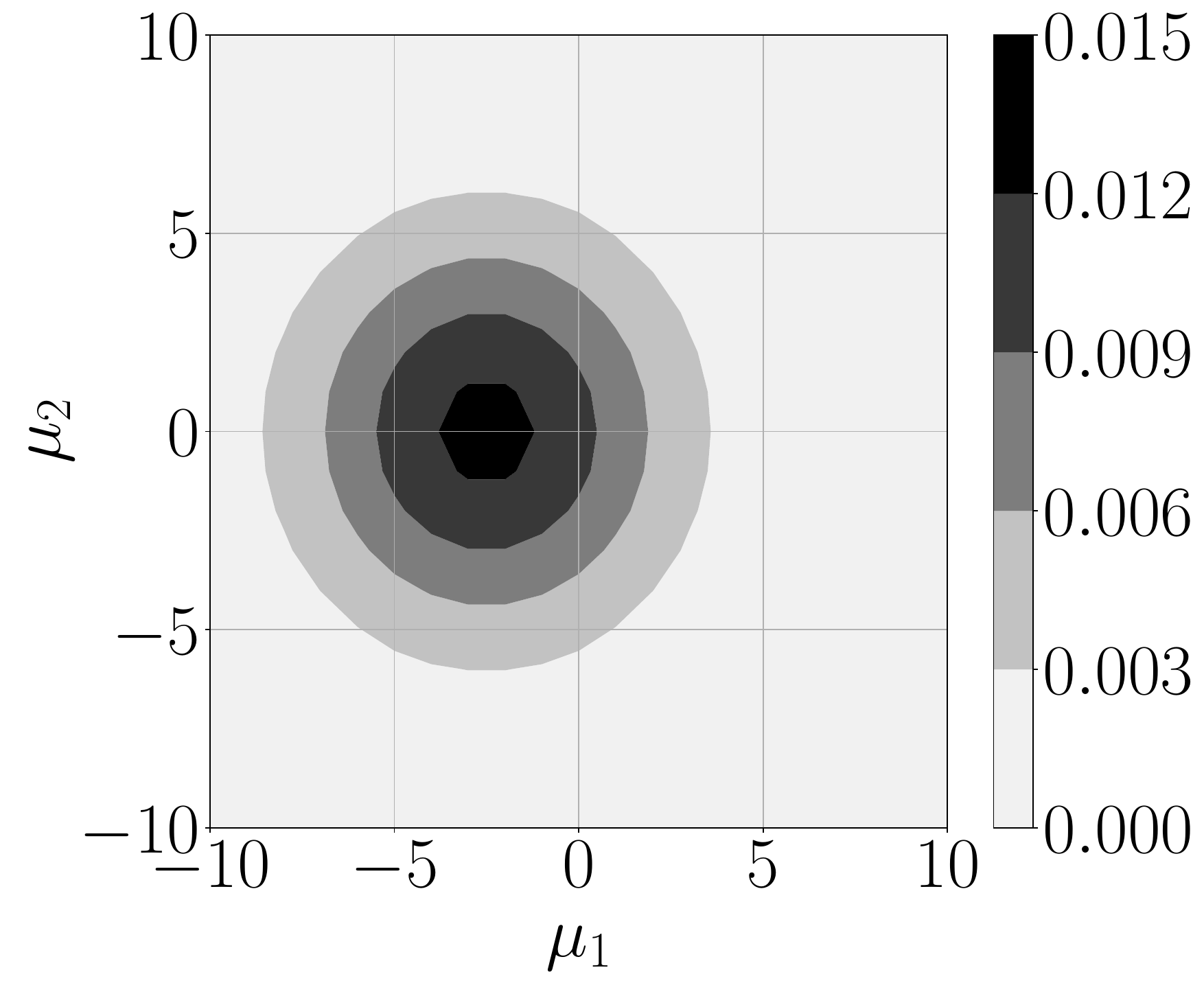}
\caption[]{$\eta_1^{(1,2)}=1,\eta_2^{(1,2)}=1$}
        \end{subfigure}
\hfill
        \begin{subfigure}[c]{.3\linewidth}
 \includegraphics[width=\linewidth]{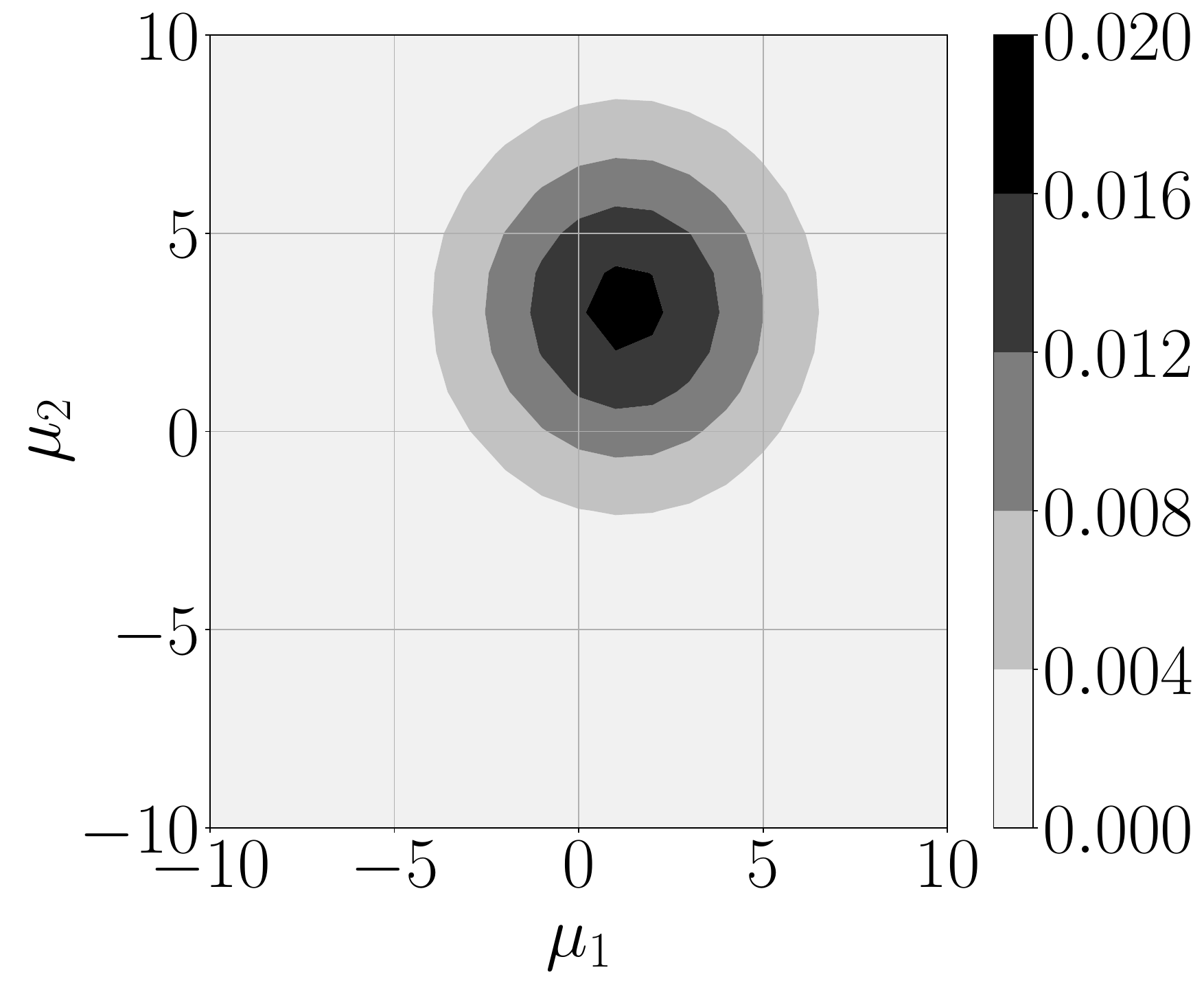}
\caption[]{$\eta_1^{(1,2)}=1,\eta_2^{(1,2)}=0.5$}
\end{subfigure}
\hfill
    \begin{subfigure}[c]{.3\linewidth}
\includegraphics[width=\linewidth]{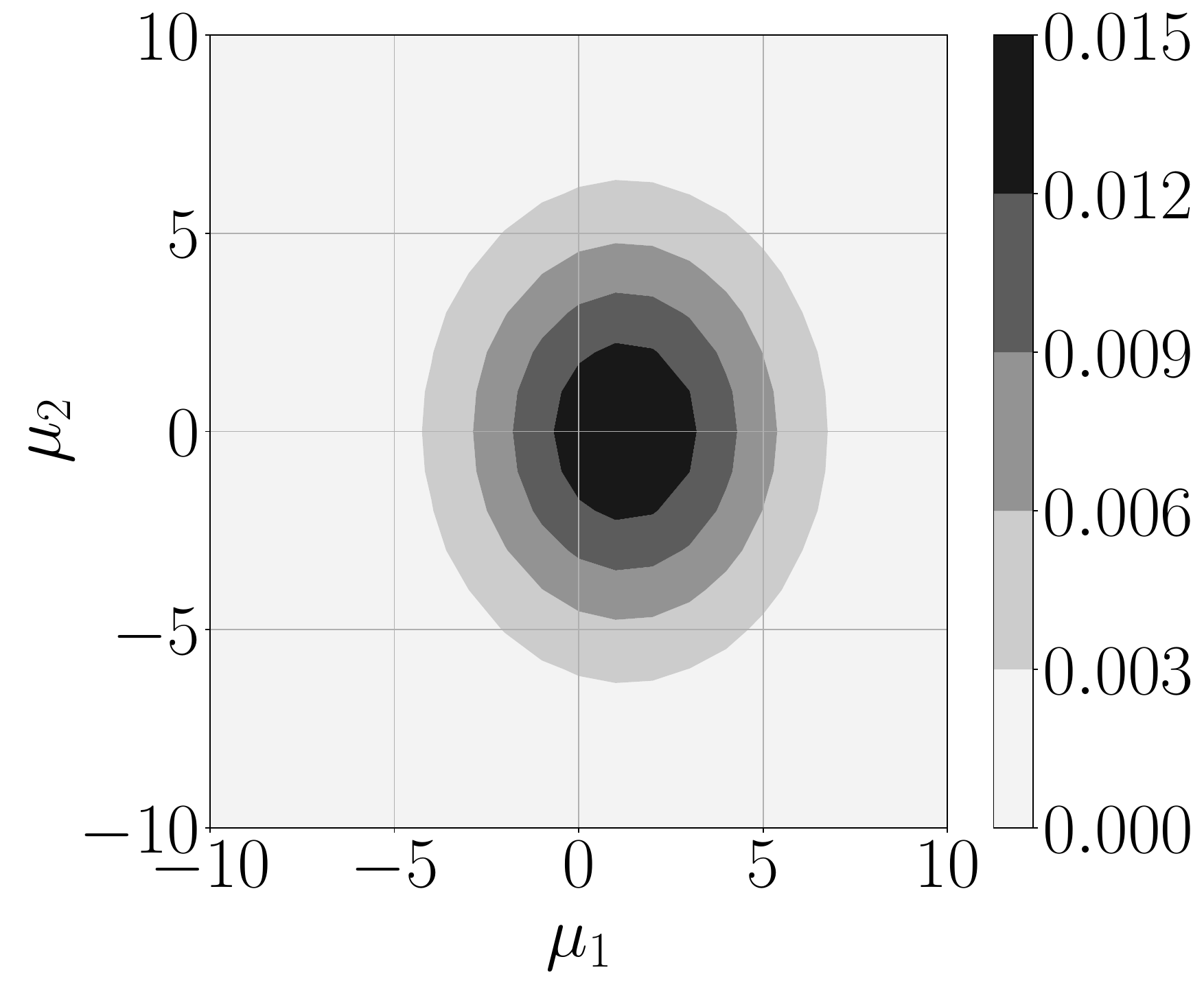}
\caption[]{$\eta_1^{(1,2)}=1,\eta_2^{(1)}=1,\eta_2^{(2)}=0.5$}
        \end{subfigure}
        \caption{Double homodyne statistical distribution of photocount differences
          computed from Eq.~\eqref{eq:P_mu1m2}
          for various detector efficiencies at $\alpha=0.5$ and $\alpha_L=5$.
          All the beam splitters are taken to be balanced.
}
\label{fig:dh-statistics}
\end{figure*}

In this section,
we consider the eight-port double homodyne scheme
depicted in Fig.~\ref{fig:double-homodyne}
(see, e.g., Refs.~\cite{Vogel:bk:2006,lahti2010realistic}).
This measurement scheme is known to allow reconstructing the
Husimi $Q$ function of the signal state~\cite{Richter:98} that provides complete information about the signal
state which may be used in CV-QKD protocols. It should be noted that restoration of the
complex amplitude of the state completely serves as the basis for composable security
proofs~\cite{PhysRevLett.93.170504,PhysRevLett.118.200501}.
%~\cite{pirandola2024improvedcomposablekeyrates,pascualgarcía2024improvedfinitesizekeyrates}.

Figure~\ref{fig:double-homodyne} shows two homodyne setups
with elements such as beam splitters BS$_i$
and photodetectors $D_{1,2}^{(i)}$ labeled with the index $i\in\{1,2\}$.
Referring to Fig.~\ref{fig:double-homodyne},
the LO and signal modes are transmitted through
the beam splitters BS$_L$ and BS$_S$, respectively.
Similarly to the analysis performed in the previous section,
we assume that the modes are in the coherent states,
$\ket{\alpha}$ and $\ket{\alpha_L}$. So, we
have the amplitudes 
\begin{align}
  &
    \label{eq:amplitudes-SL-i}
    \alpha^{(1)}=C_S\alpha,\quad \alpha^{(2)}=S_S\alpha,
    \notag
  \\
  &
    \alpha_L^{(1)}=C_L\alpha_L,\quad \alpha_L^{(2)}=-iS_L\alpha_L,
\end{align}
where $\alpha^{(i)}$ ($\alpha_{L}^{(i)}$) stands for the amplitude
describing the input coherent state of the signal (LO) mode
of the homodyne labeled by the upper index $i\in \{1,2\}$. 
Note that the phase factor $-i=e^{-i\pi/2}$ in the expression for
$\alpha^{(2)}_L$ is introduced by
a suitably chosen phase shifter placed before
the corresponding input port
of the beam splitter BS$_2$.

Direct calculation shows that
the joint statistics of the difference photocount events
is determined by the product
of the Skellam distributions given by
\begin{align}
  &
    \label{eq:P_mu1m2}
    \prob(\mu_1,\mu_2)=\prob_1(\mu_1)\prob_2(\mu_2),
    \quad \mu_i=m_1^{(i)}-m_2^{(i)},
  \\
  &
  \label{eq:Skel-i}
  \prob_i(\mu_i)=e^{-\eta_1^{(i)}|\alpha_1^{(i)}|^2}e^{-\eta_2^{(i)}|\alpha_2^{(i)}|^2}
    \Biggl(\frac{\eta_1^{(i)}|\alpha_1^{(i)}|^2}{\eta_2^{(i)}|\alpha_2^{(i)}|^2}\Biggr)^{\mu_i/2}
    \notag
  \\
  &
    \times
I_{\mu_i}\bigl(2\sqrt{\eta_1^{(i)}\eta_2^{(i)}|\alpha_1^{(i)}|^2|\alpha_2^{(i)}|^2}\bigr),
\end{align}
where, similar to the homodyne scheme,
the amplitudes of the coherent states at the output ports of the beam splitter
BS$_i$
\begin{align}
  &
    \label{eq:amplitudes-i}
    \alpha_1^{(i)}=C_i\alpha^{(i)}+S_i\alpha_L^{(i)},\quad
    \alpha_2^{(i)}=-S_i\alpha^{(i)}+C_i\alpha_L^{(i)}
\end{align}
are expressed in terms of the transmission and reflection amplitudes,
$C_i=\cos\theta_i$ and $S_i=\sin\theta_i$.

We can now apply Eqs.~\eqref{eq:Gaussian}-\eqref{eq:mu_G}
to approximate each Skellam distribution on
the right hand side of Eq.~\eqref{eq:P_mu1m2}
and derive the Gaussian approximation
for the double homodyne scheme in the form:
\begin{align}
  &
  \label{eq:Gaussian-8p}
    \prob_G(\mu_1,\mu_2)=G(\mu_1-\mu_G^{(1)};\sigma_G^{(1)})G(\mu_2-\mu_G^{(2)};\sigma_G^{(2)}),
    \\
  &
    \label{eq:sigma_G-i}
    \sigma_G^{(i)}=
    (\eta_1^{(i)}S_i^2+\eta_2^{(i)}C_i^2) |\alpha_L^{(i)}|^2,
  \\
  &
    \label{eq:mu_G-i}
    \mu_G^{(i)}=
    (\eta_1^{(i)}S_i^2-\eta_2^{(i)}C_i^2) |\alpha_L^{(i)}|^2
    +C_iS_i (\eta_1^{(i)}+\eta_2^{(i)})|\alpha_L^{(i)}| 
    \notag
  \\
  &
    \times
    2\Re\alpha^{(i)} e^{-i\phi},\quad \phi=\arg\alpha_L.
\end{align}
Similarly to Eq.~\eqref{eq:Pgood-w-def},
it is useful to put
the probability~\eqref{eq:Gaussian-8p}
into the following quadrature form:
\begin{align}
  \label{eq:P_G-x1x2}
  &
    \prob_G(x_1,x_2)\equiv \prob_G^{(\ind{DH})}(x_1,x_2)%\equiv 
    =\frac{1}{2\pi\sqrt{\sigma_G^{(1)}\sigma_G^{(2)}}}
  \notag
  \\
  &
                      \times \exp\biggl[
    -\frac{(x_1-\Re\alpha e^{-i\phi})^2}{\sigma_1}
    -\frac{(x_2-\Im\alpha e^{-i\phi})^2}{\sigma_2}
    \biggr]
\end{align}
where $x_i$ are the quadrature variables given by
\begin{align}
  \label{eq:x-1}
  x_1&=
  \frac{1}{2(\eta_1^{(1)}+\eta_2^{(1)})C_1S_1C_S}
  \notag
  \\
  &
    \times
  \left\{
  \frac{\mu_1}{|\alpha_L^{(1)}|}-(\eta_1^{(1)}S_1^2-\eta_2^{(1)}C_1^2) |\alpha_L^{(1)}|
    \right\},
  \\
  \label{eq:x-2}
  x_2&=
  \frac{1}{2(\eta_1^{(2)}+\eta_2^{(2)})C_2S_2S_S}
  \notag
  \\
  &
    \times
  \left\{
  \frac{\mu_2}{|\alpha_L^{(2)}|}-(\eta_1^{(2)}S_2^2-\eta_2^{(2)}C_2^2) |\alpha_L^{(2)}|
    \right\},
\end{align}
and relations
\begin{align}
  &
  \label{eq:sigm-i}
    \sigma_1=\frac{\sigma_x^{(1)}}{ 2 C_S^2},
    \quad
    \sigma_2=\frac{\sigma_x^{(2)}}{ 2 S_S^2},
  \\
  &
    \label{eq:sigm-x-i}
    \sigma_x^{(i)}=\frac{\eta_1^{(i)}S_i^2+\eta_2^{(i)}C_i^2}{
    \left[(\eta_1^{(i)}+\eta_2^{(i)})C_iS_i\right]^2
    }
\end{align}
give the quadrature variances $\sigma_1$ and $\sigma_2$.

As in Sec.~\ref{sec:homodyne},
formula~\eqref{eq:P_G-x1x2}
giving the $Q$ symbol of POVM (see Eq.~\eqref{eq:PG-as-avr})
provides the starting point for reconstruction of the POVM
describing the double homodyne measurements.
The covariance matrix of the POVM reads
\begin{align}
\label{eq:Sigma_m-DH}
  &
  \Sigma_m^{(\ind{DH})}=R(\phi)\diag(\delta_1,\delta_2)\tcnj{R}(\phi),
  \\
  \label{eq:sigma-i}
  &
    \delta_i=2\sigma_i-1,
\end{align}
whereas the column vector of the averaged quadratures
is given by
\begin{align}
  \label{eq:r_m-DH}
  \vc{r}_m^{(\ind{DH})}=R(\phi)
  \begin{pmatrix}
    x_1\\x_2
  \end{pmatrix}.
\end{align}

In the case, where all the beam splitters are balanced
and the photodetection is perfect,
we have
\begin{align}
  &
  \label{eq:P0-hetero}
    \prob_{\ind{CS}}^{(\ind{DH})}\left(x_1,x_2\right)=\frac{\left|\avr{\alpha|ze^{i\phi}}\right|^2}{\pi
    \left|\alpha_L\right|^2},
    \notag
  \\
  &
    \left|\avr{\alpha|z e^{i\phi}}\right|^2=e^{-|z e^{i\phi}-\alpha|^2},
    \quad
    z= x_1+ix_2,
\end{align}
so that the POVM is proportional to the projector onto the coherent state
$\hat{R}(\phi)\ket{z}=\ket{\left(x_1+ix_2\right)e^{i\phi}}$:
\begin{align}
  \label{eq:POVM-hetero-ideal}
    \hat{\Pi}_{\ind{CS}}^{(\ind{DH})}(x_1,x_2)=\frac{1}{\pi \left|\alpha_L\right|^2} \ket{z e^{i\phi}}\bra{z e^{i\phi}},
\end{align}
with the coherent state covariance matrix
\begin{align}
  \label{eq:tilde-Sigma-DH-0}
  \Sigma_{\ind{CS}}=\mathbbm{1}
\end{align}
equal to the identity matrix.

For nonideal measurements,
following the steps of the general procedure described in Sec.~\ref{subsec:hom-POVM},
we need to identify the projectors representing the ideal measurements.
At this stage, it is natural to assume these projectors
are given by the coherent state projectors
$\ket{z e^{i\phi}}\bra{z e^{i\phi}}$.
So,
the covariance matrix of
the additive noise quantum channel~\eqref{eq:Phi_N}
is given by
\begin{align}
  &
  \label{eq:Sigma-N-CS}
    \tilde{\Sigma}_N^{(\ind{DH})}=\Sigma_m^{(\ind{DH})}-\Sigma_{\ind{CS}}
    \notag
  \\
  &
  =4 R(\phi)\diag(\tilde{\sigma}_N^{(1)},\tilde{\sigma}_N^{(2)})\tcnj{R}(\phi),
  \\
  &
    \label{eq:tsigm-N-i}
    4\tilde{\sigma}_{N}^{(i)}=\delta_i-1=2(\sigma_i-1).
\end{align}

It is not difficult to check
with the help of the convolution relation~\eqref{eq:convolution}
that the probability~\eqref{eq:P_G-x1x2}
can be expressed
as a Gaussian superposition of the coherent state Husimi functions
\begin{align}
  \label{eq:P_G-x1x2-2}
    \prob_G^{(\ind{DH})}(x_1,x_2)&=\frac{\sqrt{\sigma_1\sigma_2}}{2\pi\sqrt{\sigma_G^{(1)}\sigma_G^{(2)}}}
\int\dd^2 \beta\,
    G(x_1-\beta_1;\tilde{\sigma}_N^{(1)})
    \notag
  \\
  &
    \times
    G(x_2-\beta_2;\tilde{\sigma}_N^{(2)})
    |\avr{\beta e^{i\phi}|\alpha}|^2,
\end{align}
where $\beta=\beta_1+i\beta_2$, and
$\sigma_N^{(i)}$ is
the excess noise variance  given by Eq.~\eqref{eq:tsigm-N-i}.
The corresponding expression for
POVM reads
\begin{align}
  \label{eq:povm-hetero}
    \hat{\Pi}_G^{(\ind{DH})}(x_1,x_2)&=\frac{\sqrt{\sigma_1\sigma_2}}{2\pi\sqrt{\sigma_G^{(1)}\sigma_G^{(2)}}}
\int \dd^2 \beta\,
    G(x_1-\beta_1;\tilde{\sigma}_N^{(1)})
    \notag
  \\
  &
    \times
    G(x_2-\beta_2;\tilde{\sigma}_N^{(2)})
    \ket{\beta e^{i\phi}}\bra{\beta e^{i\phi}}.
\end{align}
% with covariance matrix
% \begin{align}
%   \label{eq:tilde-Sigma-DH}
%   \tilde\Sigma^{\ind{DH}}&\equiv\ind{diag}\left(\delta_1,\delta_2\right)
%   \notag\\&=\mathbbm{1}+\ind{diag}\left(\delta_1-1,\delta_2-1\right)\notag\\
%   &\equiv \tilde\Sigma^{\ind{DH}}_0+\tilde\Sigma^{\ind{DH}}_N.
% \end{align}
An important point is that
the results given by Eq.~\eqref{eq:P_G-x1x2-2} and
Eq.~\eqref{eq:povm-hetero}
are well defined only if
the excess noise covariance matrix~\eqref{eq:Sigma-N-CS}
is positive semi-definite: $\tilde\Sigma^{\ind{(DH)}}_N\ge 0$.
The latter requires the variances
$\tilde{\sigma}_N^{(2)}$ and $\tilde{\sigma}_N^{(2)}$ to be non-negative,
which is equivalent to the inequalities:
$\sigma_1\ge 1$ and $\sigma_2\ge 1$ (see Eq.~\eqref{eq:tsigm-N-i}).
From Eq.~\eqref{eq:sigm-i},
we arrive at the conditions
$\sigma_x^{(1)}\ge 2 C_S^2$ and $\sigma_x^{(2)}\ge 2 S_S^2$ be met.

When the beam splitter BS$_S$
is balanced $2 C_S^2=2 S_S^2=1$,
the applicability of the above expression for POVM
is guaranteed by the
inequality (see Eq.~\eqref{eq:sigma-x-min})
\begin{align}
  \label{eq:sigma-x-min-i}
  \sigma_x^{(i)}\ge \left(
\frac{\sqrt{\eta_1^{(i)}}+\sqrt{\eta_2^{(i)}}}{\eta_1^{(i)}+\eta_2^{(i)}}
  \right)^2\ge 1.
\end{align}
Otherwise, either $2 C_S^2$ or $2 S_S^2$ will be above unity,
and the coherent-state representation is valid only if
the value of the corresponding variance $\sigma_x^{(i)}$ is sufficiently high.
For example, at $\eta_{1,2}^{(i)}=\eta<1/2$,
the minimal values of $\sigma_x^{(i)}$ are higher than $2$
(see Eq.~\eqref{eq:sigma-x-min-i})
and the noise variance will be positive at
any imbalance of the signal mode beam splitter
because $\max\{2 C_S^2,2 S_S^2\}\le 2$.

When one of the noise variances~\eqref{eq:tsigm-N-i} is negative,
the POVM reconstruction procedure needs to be generalized.
We shall present details on this generalization in the next section.
Meanwhile,
in the concluding part of this section,
we confine ourselves to the cases where $\sigma_N^{(i)}$ are positive.

The effects of photodetection asymmetry
are illustrated in Fig.~\ref{fig:dh-statistics}
which presents numerical results for
the double homodyne distribution~\eqref{eq:P_mu1m2}
in the photocont difference $\mu_1$-$\mu_2$ plane.
Referring to Fig.~\ref{fig:dh-statistics},
in addition to the shift of the distribution,
the asymmetry induced difference of the variances
at $\eta_1^{(1)}+\eta_2^{(1)}\ne \eta_1^{(2)}+\eta_1^{(2)}$
manifests itself as the two dimensional anisotropy
of the double homodyne distribution.

%%%%%%%%%%%%%%%%%%%%%%%%%%%%
\subsection{Positive operator-valued measure and squeezed states}
\label{subsec:gen-POVM}
%%%%%%%%%%%%%%%%%%%%%%%%%%%%%%

In this section, we apply our general procedure outlined
in
Sec.~\ref{subsec:hom-POVM} to
double homodyne measurements so as to derive the POVM expressed in
terms of the projectors describing sharp (ideal)
measurements. According to the previous section, the class of such
measurement states, which are assumed to be Gaussian, should be
extended beyond the family of the coherent states even though these
states represent the measurements in the limiting case of balanced beam splitters and perfect
photodetection.
So, we begin with  the generalized representation of
noiseless measurements
that uses a set of
pure states enlarged so as to include the squeezed states
taken in the form
% From Eq.~\eqref{eq:sigm-N-i},
% the expression for the POVM
% in the form of incoherent Gaussian superposition
% of coherent states is justified only if
% both the quadrature variances, $\sigma_1$ and $\sigma_2$,
% exceed unity.
% In this section, we show that our procedure employed
% for derivation of the double homodyne POVM can be
% suitably generalized by enlarging a set of the pure states to include
% the squeezed coherent states
\begin{align}
  \label{eq:squeezed-def}
  \ket{z,r,\phi}=\hat{R}(\phi)\hat{D}(z)\hat{S}(r)\ket{0}=
  \hat{D}(z e^{i\phi})\hat{S}(r e ^{2i\phi})\ket{0},
\end{align}
where $\hat{D}(\beta)$ ($\hat{S}(\zeta)$) is the displacement
(squeezing) operator  given by
Eq.~\eqref{eq:D-def}
(Eq.~\eqref{eq:S-def}).

Then the covariance matrix of the noiseless POVM
proportional to the projectors
$\ket{z,r,\phi}\bra{z,r,\phi}$ is given by
% By using this squeezed state distribution instead of the coherent state one given in
% Eq.~\eqref{eq:POVM-hetero-ideal}, we are immediately led to the POVM's covariance matrix in the
% form:
%Constructing POVM's~\eqref{eq:Pi_G-gen-x1x2} covariance matrix from Eq.~\eqref{eq:P_G-gen-x1x2-2} yields
\begin{align}
  \label{eq:Sigma-id-DH}
  &
  \Sigma_{\ind{id}}^{(\ind{DH})}=
    R(\phi)
    \begin{pmatrix}
      e^{2r}&0\\0&e^{-2r}
    \end{pmatrix}
    \tcnj{R}(\phi),
\end{align}
whereas formula~\eqref{eq:r_m-DH} giving
the vector of the averaged quadratures
remains applicable to the case of the squeezed states
As a result,
the non-normalized Husimi distribution
for the squeezed state~\eqref{eq:squeezed-def} takes
the form
\begin{align}
  \label{eq:squeezed-Q}
  &
    \left|\avr{z e^{i\phi},r e^{2 i \phi}|\alpha}\right|^2=\frac{1}{\cosh r}
    \exp\Biggl[-
    \frac{2}{e^{2 r}+1}
    \left(x_1-\tilde{\alpha}_1\right)^2
    \notag
  \\
  &
    -\frac{2}{e^{-2 r}+1}\left(x_2-\tilde{\alpha}_2\right)^2
      \Biggr],
  \\
  \label{eq:tbeta-alpha}
  &
  \tilde{\alpha}_1=\Re \alpha e^{-i\phi},
  \quad
  \tilde{\alpha}_2=\Im \alpha e^{-i\phi}.
\end{align}

From Eqs.~\eqref{eq:Sigma_m-DH} and~\eqref{eq:Sigma-id-DH},
we derive the covariance matrix of the additive noise channel
\begin{align}
  &
    \label{eq:Sigma-N-DH}
    \Sigma_{N}^{(\ind{DH})}(r)=
    \Sigma_{m}^{(\ind{DH})}-\Sigma_{\ind{id}}^{(\ind{DH})}
    \notag
  \\
  &
    = R(\phi)
\begin{pmatrix}
      4\sigma_N^{(1)}(r) &0\\0& 4 \sigma_N^{(2)}(r)
    \end{pmatrix}
    \tcnj{R}(\phi)\ge 0,
\end{align}
where
\begin{align}
  \label{eq:sigma_N-sq-2}
  &
    4\sigma_N^{(1)}(r)=\delta_1-e^{2r},
    \quad
    4\sigma_N^{(2)}(r)=\delta_2-e^{-2r}.
\end{align}
Since this matrix should be positive semidefinite,
the noise variances~\eqref{eq:sigma_N-sq-2}
cannot take negative values.

% \begin{align}
%   \label{eq:Pi_dh}
%   \Sigma^{\ind{DH}}&=
%     \begin{pmatrix}
%       e^{2r}&0\\0&e^{-2r}
%     \end{pmatrix}+
%     \begin{pmatrix}
%         \delta_1-e^{2r}&0\\
%         0&\delta_2-e^{-2r}
%     \end{pmatrix}\notag\\ &\equiv \Sigma_1^{\ind{DH}}+\Sigma_N^{\ind{DH}},
% \end{align}
% where, analogously to Eq.~\eqref{eq:tilde-Sigma-DH}, we defined matrices of perfect measurement
% $\Sigma_1^{\ind{DH}}$ (squeezed state covariance matrix) and excess noise $\Sigma_N^{\ind{DH}}$,
% explicitly written as
% \begin{align}
%   \label{eq:def-sigma_dh-1}
%   &\Sigma^{\ind{DH}}_1=
%     \begin{pmatrix}
%       e^{2r}&0\\0&e^{-2r}
%     \end{pmatrix},\\
%     \label{eq:def-sigma_dh-N}
%   &\Sigma^{\ind{DH}}_N=\begin{pmatrix}
%         \delta_1-e^{2r}&0\\
%         0&\delta_2-e^{-2r}
%     \end{pmatrix}.
% \end{align}
% Note that the condition~\eqref{eq:sigma_N-sq-2} is equivalent to the requirement of $\Sigma_N^{\ind{DH}}$ being positive.
%Now, we are to require that the elements of $\Sigma_N^{\ind{DH}}$ to be positive:
% To ensure physical consistency, we require the excess noise matrix $\Sigma_N^{\ind{DH}}$
% to be positive semidefinite:

% where $\delta_i\equiv2\sigma_i-1$ as previosly defined.
% These expressions present the extension of
% the relations~\eqref{eq:sigm-N-i} to the case with non-vanishing squeezing parameter.
% % As an immediate consequence of Eq.~\eqref{eq:sigma_N-sq-1}, we find that the conditions for the
% % noise variances to be positive definite can be written in the form of two inequalities

For subsequent analysis, it will be useful to express the elements $\delta_i$
that enter the covariance matrix  given by Eq.~\eqref{eq:sigma-i}
in
terms of the imbalance (reflection-to-transmission) ratio of the input beam splitter, $q$, and the
parameters $\sigma_{x}^{(1)}$ and $\sigma_{x}^{(2)}$ (see Eq.~\eqref{eq:sigm-x-i}):
\begin{subequations}
  \label{eq:delta_12}
\begin{align}
  \label{eq:delta_1}
  &
    \delta_1\equiv2\sigma_1-1=(q+1)\sigma_x^{(1)}-1\ge q=\frac{S_S^2}{C_S^2},
  \\
  &
    \label{eq:delta_2}
    \delta_2\equiv 2\sigma_2-1=(q^{-1}+1)\sigma_x^{(2)}-1\ge q^{-1}=\frac{C_S^2}{S_S^2},
\end{align}
\end{subequations}
where we used the fact that $\sigma_x^{(i)}\ge 1$ (see Eq.~\eqref{eq:sigma-x-min-i}).

\begin{figure}[t]
    \centering
    \includegraphics[width=0.9\linewidth]{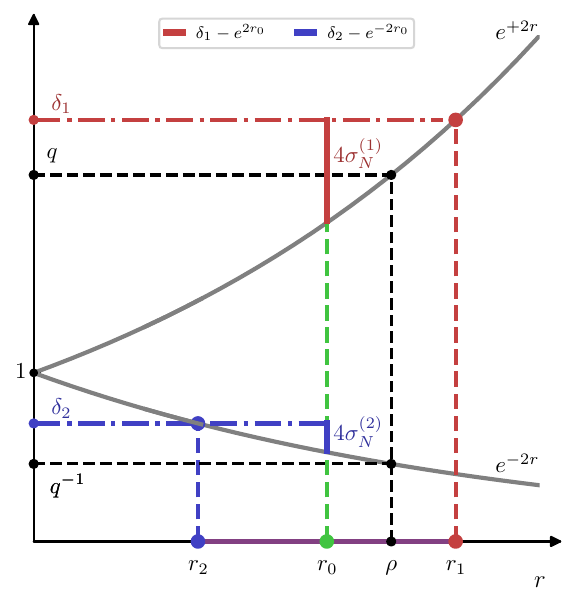}
    \caption{ Geometry in the $\delta$-$r$ (principal variance-squeezing parameter) plane.
      Noise variances~\eqref{eq:sigma_N-sq-2}
      are positive when the principal variance $\delta_1$ ($\delta_2$)
      is above $e^{2 r}$ ($e^{-2r}$), so that the squeezing parameter $r$
      is ranged between $r_1$ and $r_2$.
      Solid grey lines represent the graphs of the exponents:
      $e^{2r}$ and $e^{-2r}$.
      In the case, where $\delta_1=q$ and $\delta_2=q^{-1}$,
      the interval is reduced to the point $r_1=r_2=\rho$.}
    \label{fig:squeezing}
\end{figure}

We assume without the loss of generality
that the reflectance of the input beam splitter BS$_S$ is
larger than its transmittance,
so that the imbalance ratio is above unity 
\begin{align}
  \label{eq:q-param}
  &
    q=\frac{1-C_S^2}{C_S^2}=\frac{S_S^2}{C_S^2}\ge 1.
\end{align}
As shown in Fig.~\ref{fig:squeezing},
this implies that $\delta_1$
is above unity: $\delta_1\ge q\ge 1$,
while the minimal value of $\delta_2$
is $q^{-1}\le 1$.
It is illustrated that
the noise variances are positive,
provided that the squeezing parameter $r$
is ranged between
the endpoints of the interval given by
\begin{align}
  \label{eq:r1-r2}
  &
    r\in \left[r_2,r_1\right],\;
    r_1=\ln\delta_1^{1/2},
    \;
    r_2=\ln\delta_2^{-1/2}%\max\{\ln\delta_2^{-1/2},0\}.
\end{align}

An important point is that the value of the squeezing parameter is
generally not
uniquely determined by
the conditions~\eqref{eq:sigma_N-sq-2}.
Referring to Fig.~\ref{fig:squeezing},
the sole exception is provided by the
limiting case of perfect homodyne measurements
with $\sigma_{x}^{(1)}=\sigma_{x}^{(2)}=1$ and $\delta_{1}=\delta_{2}^{-1}=q$.
In this case, the
squeezing parameter is uniquely defined
by the imbalance ratio of the input beam splitter
\begin{align}
  &
  \label{eq:r-ideal}
  r_1=r_2=\rho=\frac{1}{2}\ln q=\ln\sqrt{\frac{1-C_S^2}{C_S^2}}
\end{align}
and the probability~\eqref{eq:P_G-x1x2}
expressed in terms of the distribution~\eqref{eq:squeezed-Q}
\begin{align}
  &
    \label{eq:P_G-ideal}
    \prob_{G}^{(1)}(x_1,x_2)=\frac{\cosh \rho}{2\pi |\alpha_L^{(1)}\alpha_L^{(2)}|}
    \left|\avr{z e^{i\phi},\rho e^{2i\phi}|\alpha}\right|^2,
\end{align}
yields the POVM
\begin{align}
    \label{eq:Pi_G-ideal}
    \hat{\Pi}_{G}^{(1)}(x_1,x_2)&=\frac{\cosh \rho}{2\pi |\alpha_L^{(1)}\alpha_L^{(2)}|}
    \notag
  \\
  &
  \times
    \ket{z e^{i\phi},\rho e^{2i\phi}}\bra{z e^{i\phi},\rho e^{2i\phi}},
\end{align}
which is proportional to the projector onto the pure squeezed state
$\ket{z e^{i\phi},\rho e^{2i\phi}}$.
The result~\eqref{eq:r-ideal}
was reported in Ref.~\cite{genoni2014general}.

When homodyne measurements are not
perfect due to unbalanced beam splitters and nonideal photodetectors,
the squeezing parameter is no longer uniquely defined.
So, in the interval~\eqref{eq:r1-r2},
we have the decomposition of
the probability~\eqref{eq:P_G-x1x2}
\begin{align}
  \label{eq:P_G-gen-x1x2-2}
  &
  \prob_G^{(\ind{DH})}(x_1,x_2)
                                 =\frac{\sqrt{\sigma_1\sigma_2}}{2\pi\sqrt{\sigma_G^{(1)}\sigma_G^{(2)}}}
                                 \int \dd^2 \beta G(x_1-\beta_1;\sigma_N^{(1)}(r))
\notag
  \\
  &
    \times
    G(x_2-\beta_2;\sigma_N^{(2)}(r))
    \left|\avr{\beta e^{i\phi},r e^{2i\phi}|\alpha}\right|^2, 
\end{align}
that varies with the squeezing parameter $r$.
Similarly,
the corresponding POVM 
\begin{align}
  \label{eq:Pi_G-gen-x1x2}
  &
    \hat{\Pi}_G^{(\ind{DH})}(x_1,x_2)=\frac{\sqrt{\sigma_1\sigma_2}}{2\pi\sqrt{\sigma_G^{(1)}\sigma_G^{(2)}}}
    \int \dd^2 \beta G(x_1-\beta_1;\sigma_N^{(1)}(r))
    \notag
  \\
  &
    \times
    G(x_2-\beta_2;\sigma_N^{(2)}(r))
    \ket{\beta e^{i\phi},r e^{2i\phi}}\bra{\beta e^{i\phi},r e^{2i\phi}},
\end{align}
decomposed into
the Gaussian incoherent superposition
of the pure squeezed states explicitly depends on the value of $r$.% Its' covariance matrix is given by Eq.~\eqref{eq:Pi_dh}.

Our analysis suggests that  the ambiguity (non-uniqueness) of
the Gaussian representation~\eqref{eq:Pi_G-gen-x1x2} for the double-homodyne
POVM is a universal feature that comes into play in the presence of imperfections.
Even when $\delta_2\ge 1$ and
the coherent-state representation~\eqref{eq:povm-hetero} for the POVM is well-defined,
the photocount statistics can be reproduced using
the squeezed-state representation~\eqref{eq:Pi_G-gen-x1x2}
with $r\neq 0$.
Analogously, the density matrices of mixed Gaussian states can be
expressed as randomly displaced squeezed states, and, in general, this
representation is non-unique.
In the next section, we demonstrate that
the non-uniqueness of additive noise channel
should be taken into account in
certain scenarios of security analysis of
continuous variable quantum key distribution protocols.

% One way to make the Gaussian POVM decomposition unique is to impose
% additional constraints on the noise variances, which in turn fix the
% value of the squeezing parameter.  For example, the product of noise
% variances is maximized by the midpoint of the interval:
% \begin{align}
%   \label{eq:mid-r}
%   r_0&\equiv\argmax_r\left[\sigma_N^{(1)}(r) \times \sigma_N^{(2)}(r)\right]\notag\\
%   &=\frac14\ln\frac{\delta_1}{\delta_2}=\frac{r_1+r_2}2,
% \end{align}
% which is easily verified by finding the extremum using definitions from Eq.~\eqref{eq:sigma_N-sq-2}.

% the squeezing parameter would take the minimal (maximal) value: $r=\min\{r_1,r_2\}$ ($r=\max\{r_1,r_2\}$)
%provided the constraint requires minimization of
%the noise variance $\sigma_N^{(2)}$ ($\sigma_N^{(1)}$).
%%%%%%%%%%%%%%%%%%%%%%%%%%%%%

%%%%%%%%%%%%%%%%%%%%%The non-uniqueness of additive noise channel
% should be taken into account when analyzing 
%%%%%%%%%%%%%%%%%%%%%%%%%%%%%%%%%%%%%%%%
\section{Measurement imperfections and security
of continuous variable quantum key distribution protocols}
\label{sec:protocol}
%%%%%%%%%%%%%%%%%%%%%%%%%%%%%%%%%%%%%%%%%%%%%%%%%%%%

In this section, we apply the developed formalism for asymmetric noisy
Gaussian measurements to study
how the measurement noise
manifests itself in analysis
of
continuous variable quantum key distribution
(CV-QKD) protocols
(see, e.g., Ref.~\cite{Zhang:apr:2024} for a recent review).
Specifically,
we concentrate on the
Gaussian-modulated coherent-state
(GMCS, or GG02)
CV-QKD protocol suggested in~\cite{PhysRevLett.88.057902}
and examine
how the measurement asymmetry affects its asymptotic security
in the untrusted-noise scenario~\cite{Laudenbach:aqt:2018}.

% we evaluate the mutual information, Holevo information, and asymptotic secret fraction (secure key rate
% per symbol) under the untrusted-noise scenario to quantify 
% the protocol security.

%%%%%%%%%%%%%%%%%%%%%%%%%
\subsection{Untrusted-noise scenario}
\label{subsec:untr-noise}
%%%%%%%%%%%%%%%%%%%%%%%%%

In the GMCS protocol,
the sender (Alice) prepares an ensemble of normally distributed coherent states
$\ket{\alpha=\alpha_1+i \alpha_2}$
with the Gaussian probability density
$p_{\ind{A}}(\alpha)=G(\alpha_1;V_A)G(\alpha_2;V_A)$,
so that the density matrix of the input state
\begin{align}
  \label{eq:ensemble}
  &
    \hat{\rho}_{\ind{A}}=\int\mathop{\dd^2\alpha} p_{\ind{A}}(\alpha)
    \ket{\alpha}\bra{\alpha}=\Phi_A(\ket{0}\bra{0}),
    \notag
\\
    %\label{eq:probabilities}
    &p_{\ind{A}}(\alpha)=\frac{1}{\pi V_{\ind{A}}}\exp\left[
    -\frac{|\alpha|^2}{2V_{\ind{A}}}
    \right]
\end{align}
represents randomly displaced states generated by
applying the additive noise channel $\Phi_A$ with the covariance matrix
$\Sigma_A= 4 V_A\mathbbm{1}$ to the vacuum state $\ket{0}\bra{0}$.
The states prepared by Alice are then transmitted to
the receiver (Bob)
through the quantum channel
$\mathcal{N}_{A\to B}$ which is assumed to be a composition
of two Gaussian channels~\cite{Laudenbach:aqt:2018}
given by
\begin{align}
  \label{eq:AB-channel}
  \mathcal{N}_{A\to B}=\Phi_{\xi}\circ\mathcal{E}_T,
\end{align}
where
$\mathcal{E}_T$ is the pure-loss channel with transmittance $T$
and $\Phi_{\xi}$ is the additive noise channel  with
noise covariance matrix $\Sigma_{\xi}=\xi \mathbbm{1}$.
Channel noise with variance $\xi$ in this model accounts for various
stochastically independent noise sources, such electronic noise,
non-infinite CMRR, and relative intensity noise
(RIN)~\cite{Laudenbach:aqt:2018}.
In order to decode the signal,
Bob measures the output state of the quantum channel~\eqref{eq:AB-channel}.

This is  a prepare-and-measure (PM) protocol,
where
the mutual information between Alice and Bob
is determined by
the joint probability distribution of
the measurement outcomes and the components of the coherent state
amplitude.
It is not difficult to show
(see Appendix~\ref{sec:MI-and-EBtoPm} for technical details)
that, when Bob performs homodyne detection,
the mutual information is given by
\begin{align}
  \label{eq:I_H_res}
  I_{\ind{AB}}^{(\ind{H})}=\frac{1}{2}\log_2\left[1+\frac{4TV_{\ind{A}}}{\sigma_x+\xi}\right],
\end{align}
where $\sigma_x$ is given by Eq.~\eqref{eq:sigma-x},
whereas, if Bob uses double-homodyne detection, the expression for
mutual information takes the form
\begin{align}
  \label{eq:I_DH_res}
  I_{\ind{AB}}^{(\ind{DH})}=\frac{1}{2}\sum_i\log_2\left[1+\frac{4TV_{\ind{A}}}
    {2 \sigma_i+\xi}\right],
\end{align}
where $\sigma_i$ are given by Eq.~\eqref{eq:sigm-x-i}.

 %This additional noise effectively enhances the quantum channel
       %parameters, thereby increasing the information accessible to
       %Eve. %Consequently, this noise contribution effectively
       %augments the quantum-channel parameters, increasing Eve’s
       %accessible information.

According to Eqs.~\eqref{eq:I_H_res}
and~\eqref{eq:I_DH_res},
the mutual information depends on the type of measurement
performed by Bob.
This information
enters the Devetak-Winter formula~\cite{Devetak:prsa:2005}
for the asymptotic secret fraction
(the number of secret bits that can be distilled per transmission
of the signal)
\begin{equation}
\label{eq:r}
K^{(\gamma)}=\beta I_{AB}^{(\gamma)}-\chi_{EB}^{(\gamma)},
\quad \gamma\in\{\ind{H},\ind{DH}\},
\end{equation}
where $0 <\beta< 1$ is the reconciliation efficiency
and $\chi_{EB}^{(\gamma)}$ is
the Holevo information.
This formula suggests using
the entanglement-based (EB) version of
the protocol, where Alice prepares a two-mode squeezed vacuum (TMSV)
state, measures both quadratures of the first mode and
sends the second mode to Bob.
As discussed in Appendix~\ref{sec:MI-and-EBtoPm} after
Eq.~\eqref{eq:I_dh},
the EB representation is equivalent to the PM protocol
because both versions 
are indistinguishable from the perspective
of the receiver and the eavesdropper (Eve).

In this picture,
the covariance matrix of
the bipartite quantum state shared by Alice and Bob,
$\hat{\rho}_{\ind{AB}}$,
can be obtained
from the TMSV covariance matrix
with Bob's mode
transmitted through the noisy channel
as follows
(see Ref.~\cite{Laudenbach:aqt:2018})
\begin{align}
  \label{eq:TMSVS-to-AB}
  \mathcal{I}_A&\otimes \Phi_{\xi}\circ\mathcal{E}_T:
\quad
    \Sigma_{\ind{TMSV}}
  \notag
  \\
  &
  =
  \begin{pmatrix}
        V\mathbbm{1}&\sqrt{V^2-1}\sigma_z\\
        \sqrt{V^2-1}\sigma_z&V\mathbbm{1}
    \end{pmatrix}
  \mapsto
    \Sigma_{\ind{AB}}
    \notag
  \\
  &
    =
    \begin{pmatrix}
        V\mathbbm{1}&c\sigma_z\\
        c\sigma_z&V_{\ind{B}}\mathbbm{1}
    \end{pmatrix}, \quad \sigma_z=\ind{diag}(1,-1),
\end{align}
where
$\mathcal{I}_A$ is the identity channel that acts on the subsystem $A$;
$\mathbbm{1}=\ind{diag}(1,1)$ is the unity matrix;
and the parameters of the covariance matrix
$\Sigma_{\ind{AB}}$ are given by
\begin{align}
  \label{eq:V-c}
  &
    V=1+4V_{\ind{A}},
    \quad c=\sqrt{T(V^2-1)},
  \\
  &
  \label{eq:V_B}
    V_{\ind{B}}=T(V-1)+1+\xi.
\end{align}
% Note that Bob's variance $V_{\ind{B}}$ is exactly the same as Bob's variance calculated in PM
% protocol per Eq.~\eqref{eq:sigma-hom-def} after substituting $\sigma_x=1$, as is required to
% equate both protocols.  Measurement noise modifies this covariance matrix as

The standard approach to
the evaluation of Holevo information
is to assume that Eve holds a purification
of $\hat{\rho}_{AB}$
and use the extremality property of Gaussian
states~\cite{Miguel:prl:2006,Cerf:prl:2006}
to estimate the upper bound
of the Holevo information
from the covariance matrix
$\Sigma_{\ind{AB}}$.

In the untrusted-noise scenario,
where additional noise
responsible for detection imperfections
is assumed to be accessible to Eve,
the density matrix $\hat{\rho}_{\ind{AB}}$ should be modified.
In our case, the noisy POVMs
(see Eqs.~\eqref{eq:homodyne-povm}
and~\eqref{eq:Pi_G-gen-x1x2})
are generated from the POVMs of ideal measurements
by applying the dual of a Gaussian quantum channel
of the following form
\begin{align}
  \label{eq:Phi_N_Pi}
  \hat{\Pi}_G^{(\gamma)}=
  \cnj{[\Phi_N^{(\gamma)}]}(\hat{\Pi}_{\ind{id}}^{(\gamma)})=
  \Phi_N^{(\gamma)}(\hat{\Pi}_{\ind{id}}^{(\gamma)}),
\end{align}
where $\Phi_N^{(\gamma)}$
is the additive noise channel
with the covariance matrix
$\Sigma_N^{(\gamma)}$
(Eqs.~\eqref{eq:Sigma-N-H} and~\eqref{eq:Sigma-N-DH}
give
$\Sigma_N^{(\ind{H})}$ and $\Sigma_N^{(\ind{DH})}$,
respectively).
When this channel is under the control of the eavesdropper, 
Eve holds a purification of the state
$\hat{\rho}_{\ind{AB}}^{(\gamma)}=\mathcal{I}_A\otimes\Phi_N^{(\gamma)}(\hat{\rho}_{\ind{AB}})$
with the covariance matrix given by
\begin{align}
  &
  \label{eq:sigma_m_cov}
\mathcal{I}_A\otimes \Phi_{N}^{(\gamma)}:
  \Sigma_{\ind{AB}}\mapsto
  \Sigma_{\ind{AB}}^{(\gamma)}
  \notag
  \\
  &
  =\begin{pmatrix}
    V\mathbbm{1}&c\sigma_z\\
    c\sigma_z&V_{\ind{B}}\mathbbm{1}+\Sigma_N^{(\gamma)}
\end{pmatrix},\quad \gamma\in\{\ind{H},\ind{DH}\}
\end{align}
and the measurements performed by the receiver
are assumed to be noiseless.
This is the scenario of our primary concern.
In this scenario, 
the difference between the joint and conditional entropies
\begin{align}
  \label{eq:chi}
  &
  \chi_{\ind{EB}} ^{(\gamma)}
    =
    S_{\ind{E}}-S_{\ind{E}|\ind{B}}=S_{\ind{AB}}^{(\gamma)}-S_{\ind{A}|\ind{B}}^{(\gamma)}
\end{align}
gives the Holevo information.
The joint entropy
\begin{align}
  &
  \label{eq:S_AB-gamma}
  S_{\ind{AB}}^{(\gamma)}=g(\nu_1^{(\gamma)})+g(\nu_2^{(\gamma)}),
  \\
  &
\label{eq:g(nu)}
    g(\nu)=\frac{\nu+1}{2}\log_2 \frac{\nu+1}{2}-
    \frac{\nu-1}{2}\log_2\frac{\nu-1}{2}    
\end{align}
then can be expressed in terms of
the symplectic eigenvalues of the covariance matrix~\eqref{eq:sigma_m_cov}
given by
\begin{align}
  &
    \label{eq:nu-12-gamma}
\nu_{1,2}^{(\gamma)}
    =\sqrt{
\Delta_{\gamma}/2\pm \sqrt{\Delta_{\gamma}^2/4-D_{\gamma}}
    },
    \quad
  D_{\gamma}=\det \Sigma_{\ind{AB}}^{(\gamma)},  
    \\
  &
  \label{eq:D-gamma}
  \Delta_{\gamma}=\det(V_{\ind{B}}\mathbbm{1}+\Sigma_N^{(\gamma)}) +
  V^2- 2 c^2.
\end{align}

By applying the partial measurement formula~\cite{Serafini:bk:2023}
to the covariance matrix $\Sigma_{\ind{AB}}^{(\gamma)}$,
we also obtain the conditional covariance matrix
\begin{align}
  \label{eq:serf-a|b}
  \Sigma_{\ind{A|B}}^{(\gamma)}=
  V\mathbbm{1}-c^2\sigma_z(V_{\ind{B}}\mathbbm{1}+\Sigma^{(\gamma)}_m)^{-1}\sigma_z,
\end{align}
so that the conditional entropy $S_{\ind{E|B}}=S_{\ind{A|B}}^{(\gamma)}$
reads
\begin{align}
  &
    \label{eq:S_A1B-gamma}
    S_{\ind{A|B}}^{(\gamma)}=g(\nu_3^{(\gamma)}),\quad
    \nu_3^{(\ind{\gamma})}=\sqrt{\det\Sigma_{\ind{A|B}}^{(\gamma)}},
\end{align}
where $\nu_3^{(\ind{\gamma})}$ is the simplectic eigenvalue
of $\Sigma_{\ind{A|B}}^{(\gamma)}$.
For homodyne detection,
we have
\begin{align}
  \label{eq:nu3-H}
    \nu_3^{(\ind{H})}=\sqrt{V\left(V-\frac{c^2}{V_{\ind{B}}+\sigma_N}\right)},
\end{align}
whereas,
for double homodyne detection,
the result is
\begin{align}
    \label{eq:nu3-DH}
  &
    \nu_3^{(\ind{DH})}=V
    \sqrt{\frac{\left(V_{\ind{B}}  + \delta_1 - \frac{c^{2}}V\right) \left( V_{\ind{B}}  + \delta_2
    - \frac{c^{2}}V\right)}{
    \left( V_{\ind{B}} + \delta_1\right) \left(V_{\ind{B}} + \delta_2\right)}},
\end{align}
where $\delta_i=2\sigma_i-1$ is given by Eq.~\eqref{eq:delta_12}.

\begin{figure*}[!htp]
    \centering
    \begin{subfigure}[c]{.3\linewidth}
\includegraphics[width=\linewidth, ]{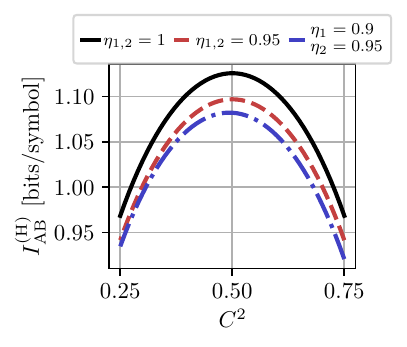}
\caption[]{}%{$I_{\ind{AB}}$}
\label{subfig:hom-a}
        \end{subfigure}
\hfill
        \begin{subfigure}[c]{.3\linewidth}
 \includegraphics[width=\linewidth, ]{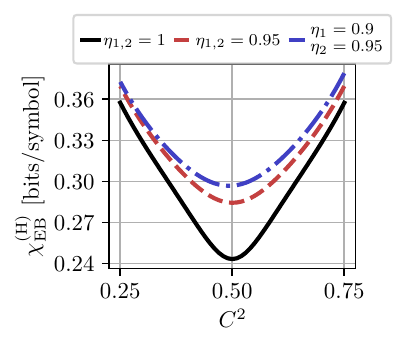}
\caption[]{}%{$\chi_{\ind{EB}}$}
\label{subfig:hom-b}
\end{subfigure}
\hfill
    \begin{subfigure}[c]{.3\linewidth}
\includegraphics[width=\linewidth, ]{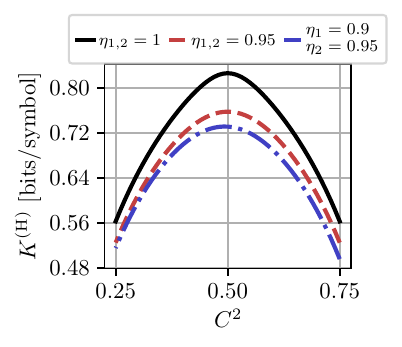}
\caption[]{}%{$R$}
\label{subfig:hom-c}
        \end{subfigure}
        \caption{
          Informational quantities computed for the case of homodyne receiver (see Fig.~\ref{fig:homodyne}).
          (a)~Mutual information, $I_{\ind{AB}}^{(\ind{H})}$, (b)~Holevo
        information, $\chi_{\ind{EB}}^{(\ind{H})}$ and (c)~asymptotic
        secret fraction, $K^{(\ind{H})}$,
        as a function of the beam splitter
        transmission, $C^2$,
        computed at different values of the detector efficiencies. The
        parameters are:
        $V=5$;
        $T=0.95$;
        $\xi=10^{-2}$; and
        $\beta=0.95$.
      }
\label{fig:homodyne-all}
\end{figure*}

%%%%%%%%%%%%%%%%%%%%%%%%%%%%%
\subsection{Numerical results}
\label{subsec:numerical-results}
%%%%%%%%%%%%%%%%%%%%%%%%%

Our analytical results for the mutual information (see Eqs.~\eqref{eq:I_H_res}
and~\eqref{eq:I_DH_res}) and the entropies (see Eqs.~\eqref{eq:S_AB-gamma} and~\eqref{eq:S_A1B-gamma})
that enter the expression for the Holevo information~\eqref{eq:chi} can now be used to quantify the
asymmetry and imbalance induced effects of noisy measurements. One of the key points is that, in the
untrusted-noise scenario, the Holevo information explicitly depends on the dual channel
representation of noisy POVMs given by Eq.~\eqref{eq:Phi_N_Pi}.

In the case of homodyne detection, the ideal measurements are described by the quadrature eigenstate
projectors and the additive noise channel representation is uniquely determined by the excess noise
covariance matrix~\eqref{eq:Sigma-N-H}.
Figure~\ref{fig:homodyne-all} shows how the mutual information, the Holevo
information and the secret key fraction depend on the transmittance of the beam splitter used in the
homodyne detection scheme (see Fig.~\ref{fig:homodyne}).
It is seen that, in general, the beam splitter imbalance has a
detrimental effect on performance of the protocol.
Interestingly,
in the presence of asymmetry in the efficiencies of the photodetectors,
both the mutual information and secret
fraction can be maximized by introducing
a suitably unbalanced beam splitter.
A similar remark applies to minimization of the Holevo information.
Qualitatively, behavior of the curves presented
in Fig.~\ref{subfig:hom-c} agrees with
the corresponding results reported in Ref.~\cite{ruiz2023effects}
(see Fig.~6c in~\cite{ruiz2023effects}).

\begin{figure}
    \centering
\includegraphics[width=\linewidth]{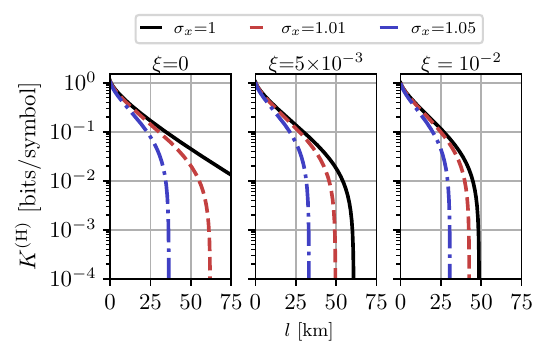}
\caption{ Asymptotic secret fraction, $K^{(\ind{H})}$, for a homodyne receiver as a function of the channel length, $l$,
  computed for various values of
  the quadrature variance given by Eq.~\eqref{eq:sigma-x}, $\sigma_x\in\{1,1.01,1.05\}$, and the excess noise,
  $\xi\in\{0,0.005,0.01\}$.
  The losses are assumed to be $20$ dB per $100$ km.  The parameters are:
  $V=5$ and $\beta=0.95$.  }
      \label{fig:K-homodyne}
      \end{figure}

\begin{figure*}[!htp]
  \centering
  \begin{subfigure}[c]{.45\textwidth}
 \includegraphics[width=\linewidth]{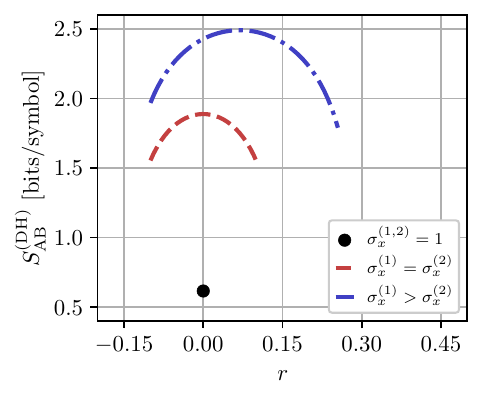}
 \caption[]{$q=1$}
\label{subfig:Entropy-a}
\end{subfigure}
\hfill
\begin{subfigure}[c]{.45\textwidth}
 \includegraphics[width=\linewidth]{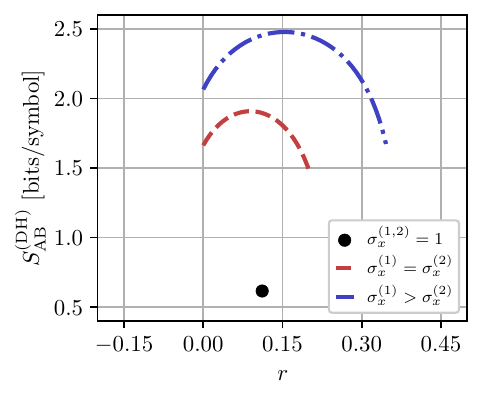}
 \caption[]{$q=1.25$}
\label{subfig:Entropy-b}
\end{subfigure}
\caption{Joint entropy, $S_{\ind{AB}}^{(\ind{DH})}$, as a function of the squeezing parameter $r$
  (see Eq.~\eqref{eq:r1-r2}) at different values of the quadrature variances
  $\sigma_x^{(1,2)}$~\eqref{eq:sigm-x-i} for (a)~the balanced input beam splitter with the imbalance
  ratio~\eqref{eq:q-param}, $q$, equal to unity and (b)~the unbalanced input beam splitter with
  $q=1.25$. Values of $\sigma_x^{(1,2)}$ are: $\sigma_x^{(1,2)}=1$ for black dot,
  $\sigma_x^{(1,2)}=1.11$ for red dashed line, and $\sigma_x^{(1)}=1.33, \sigma_x^{(2)}=1.11$ for
  blue dashdotted line.  The channel parameters are: $V=5$; $T=0.95$; and $\xi=10^{-2}$.  }
  \label{fig:Entropy}
\end{figure*}

\begin{figure}[b]
    \centering
\includegraphics[width=.7\linewidth]{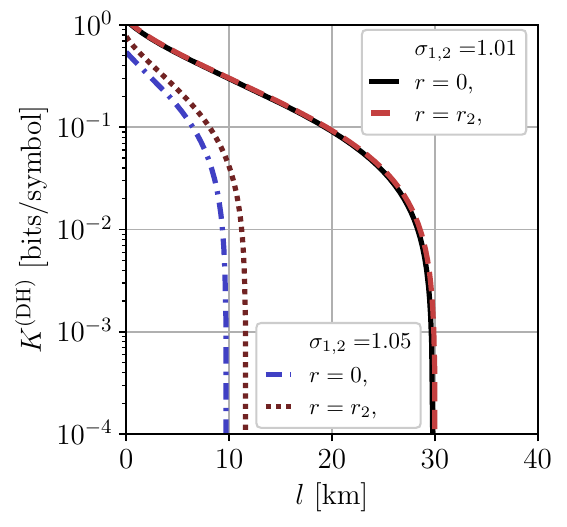}
\caption{Asymptotic secret fraction, $K^{(\ind{DH})}$,  for a double-homodyne receiver, as a function of the  channel length, $l$,
  computed for various values of
  the quadrature variance, $\sigma_{1}=\sigma_2\in\{1.01,1.05\}$ (see Eq.~\eqref{eq:sigm-i}), and the squeezing
  parameter~\eqref{eq:r1-r2}, $r\in\{0,r_2\}$. The losses are $20$ dB per $100$ km.  The
  parameters are: $V=5$; $\beta=0.95$; $\xi=10^{-2}$.}
      \label{fig:K-DH-r}
      \end{figure}

\begin{figure*}
    \centering
        \begin{subfigure}[c]{.45\linewidth}
\includegraphics[width=\linewidth]{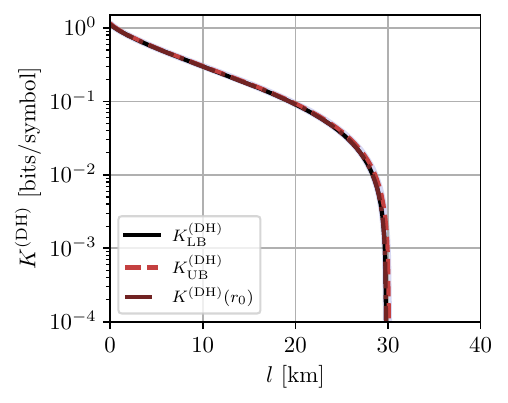}
\caption[]{$q=1$}
\label{subfig:K-vs-L-q-a}
        \end{subfigure}
        \hfill
        \begin{subfigure}[c]{.45\linewidth}
\includegraphics[width=\linewidth]{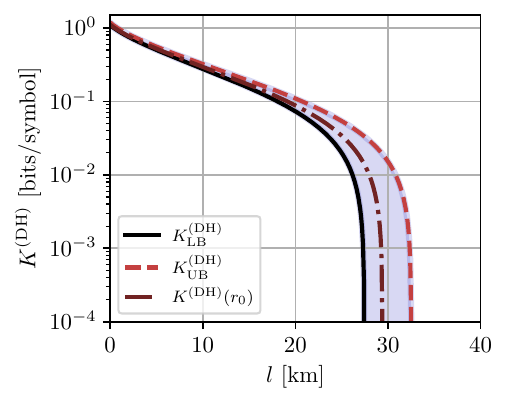}
\caption[]{$q=1.25$}
\label{subfig:K-vs-L-q-b}
        \end{subfigure}
        \caption{Asymptotic secret fraction for double-homodyne receiver, $K^{(\ind{DH})}$, as a function of the channel length,
          $l$,
          computed at $\sigma_x^{(1)}=\sigma_x^{(2)}=1.01$ (see Eq.\eqref{eq:sigm-x-i}) for (a)~the balanced input beam
          splitter with the imbalance ratio~\eqref{eq:q-param}, $q$, equal to unity and (b)~the
          unbalanced input beam splitter with $q = 1.25$. Black solid and red dashed lines
          represent the curves for the lower and upper bounds of $K^{(\ind{DH})}$:
          $K^{(\ind{DH})}_{\ind{LB}}$ and $K^{(\ind{DH})}_{\ind{UB}}$, respectively.
It is illustrated that
          a curve for $K^{(\ind{DH})}(l)$ calculated at $r_2\le r\le r_1$ lies  within the while blue shaded
          area bounded by $K^{(\ind{DH})}_{\ind{LB}}$ and $K^{(\ind{DH})}_{\ind{UB}}$.
          The losses are $20$ dB per
          $100$ km.  The parameters are: $r=r_0=(r_2+r_1)/2$; $V=5$; $\beta=0.95$; $\xi=10^{-2}$.}
        \label{fig:K-vs-L-q}
\end{figure*}

Figure~\ref{fig:K-homodyne} illustrates
how imperfect homodyne detection influence
dependence of
the asymptotic secret key fraction,
$K^{(\ind{H})}$,
on the channel length, $l$,
evaluated in the untrusted-noise scenario.
The curves presented in Fig.~\ref{fig:K-homodyne}
are computed for the quadrature variances $\sigma_x$~\eqref{eq:sigma-x}
that can be conveniently characterized by
the corresponding values of effective detection efficiencies:
$\eta_x=1/\sigma_x\in\{1,0.99,0.95\}$.
The values of the excess noise, $\xi$,
indicated in Fig.~\ref{fig:K-homodyne} are taken from
Ref.~\cite{Laudenbach:aqt:2018}.
It is shown that an increase in
the quadrature variance and
the excess noise produces similar effects
leading to a reduction in
the reach.
The latter is the maximum transmission length
at which the key rate is positive. 
Note that
the curves illustrating
the excess noise induced effects
are in good agreement with
the results
presented in Fig.~3 of
Ref.~\cite{Laudenbach:aqt:2018}.
Degradation of the perfomance in asymmetric homodyne
is also in
  a qualitative agreement with the results recently reported in Ref.~\cite{Bartlett:25} (see
  Fig.~7).

By contrast to homodyne detection,
the channel representation
for the POVM of noisy double homodyne measurements~\eqref{eq:Pi_G-gen-x1x2}
appears to depend on the properties
of the pure squeezed states~\eqref{eq:squeezed-def}
describing the noiseless POVMs.
According to Eq.~\eqref{eq:Sigma-N-DH},
the squeezing parameter, $r$,
which lies in the interval~\eqref{eq:r1-r2}
explicitly enters
the expression for the excess noise covariance matrix, $\Sigma_N^{(\ind{DH})}$.
The latter makes the covariance matrix~\eqref{eq:sigma_m_cov},
$\Sigma_{\ind{AB}}^{(\ind{DH})}$,
and the joint (Eve's) entropy~\eqref{eq:S_AB-gamma},
$S_{\ind{AB}}^{(\ind{DH})}$, $r$-dependent.
Hence, it turned out that the Holevo information
is a function of $r$.

% Note that in the case of double homodyne detection, joint entropy
% $S_{\ind{AB}}$ and thereby Holevo information $\chi_{\ind{EB}}$ are
% explicitly dependent on squeezing parameter
% $r\in[r_2,r_1]$~\eqref{eq:r1-r2}, as seen from the expression for
% covariance matrix~\eqref{eq:sigma_m_cov}.

Figure~\ref{fig:Entropy}
illustrates the dependence of the joint (Eve's) entropy,
$S_{\ind{AB}}^{(\ind{DH})}$,
on the squeezing parameter~\eqref{eq:r1-r2} (see Fig.~\ref{fig:squeezing}). 
Referring to Fig.~\ref{fig:Entropy},
in the limiting case of ideal double homodyne with
$\sigma_x^{(1)}=\sigma_x^{(2)}=1$,
the only allowed value
of $r$, $r=\rho$, is given by Eq.~\eqref{eq:r-ideal}
and the corresponding value of entropy is marked by the black dot.
Otherwise, the endpoints of the interval~\eqref{eq:r1-r2} differ.

When the input (signal) beam splitter is balanced and
the imbalance ratio~\eqref{eq:q-param} equals unity,
$q=1$,
all curves shown in Fig.~\ref{subfig:Entropy-a}
contain the zero-squeezing point with $r=0$
representing the coherent-state limiting case~\eqref{eq:povm-hetero}.
It can be seen that the maximum of the entropy
is located at this point,
provided the homodyne detection variances
are identical to
$\sigma_x^{(1)}=\sigma_x^{(2)}$.
However, this is  no longer the case when the symmetry is broken
and $\sigma_x^{(1)}\ne\sigma_x^{(2)}$.
The curves depicted in
Fig.~\ref{subfig:Entropy-b}
demonstrate that, at $q\ne 1$, the zero-squeezing point
can be outside the squeezing parameter interval~\eqref{eq:r1-r2}.
This is the case where the entropy $S_{\ind{AB}}^{(\ind{DH})}$
is ill-defined at $r=0$
due to the presence of symplectic eigenvalues smaller than unity.

Since the entropy $S_{\ind{AB}}^{(\ind{DH})}$ depends on the squeezing parameter,
this parameter will also affect both
the Holevo information $\chi_{\ind{EB}}^{(\ind{DH})}$ (see Eq.~\eqref{eq:chi})
and the asymptotic secret fraction (see Eq.~\eqref{eq:r}).
By assuming that
$\sigma_{1}=\sigma_2\in\{1.01,1.05\}$ and $q=1$, in Fig.~\ref{fig:K-DH-r}, we plot
the asymptotic secret fraction against the channel length
at $r=0$ and $r=r_2$.
It can be seen that,
at nearly sharp measurements with
$\eta_x=1/\sigma_{x}^{(i)}\approx 0.99$ and $\sigma_{1}=\sigma_2=1.01$,
the difference between the curves is negligibly  small,
whereas, at $\eta_x\approx 0.95$ and $\sigma_{1}=\sigma_2=1.05$,
the reach for the curve computed at
the coherent-state point ($r=0$),
where the entropy $S_{\ind{AB}}^{(\ind{DH})}$ takes the maximal value
(see Fig.~\ref{fig:Entropy}),
is noticeably shorter as compared to the case,
where $r=r_2$ and $S_{\ind{AB}}^{(\ind{DH})}$ is minimal.

In general, the effect of the squeezing parameter
can be conveniently described in terms of
the upper and lower bounds,
$K^{(\ind{DH})}_{\ind{UB}}$
and
$K^{(\ind{DH})}_{\ind{LB}}$,
for the asymptotic secret fraction given by the relation
\begin{align}
  &
  \label{eq:K_LU}
  K^{(\ind{DH})}_{\ind{LB}}(l)=\min_{r\in[r_2,r_1]} K^{(\ind{DH})}(l,r)\le K^{(\ind{DH})}(l,r)
  \notag
  \\
  &
  \le K^{(\ind{DH})}_{\ind{UB}}(l)=\max_{r\in[r_2,r_1]} K^{(\ind{DH})}(l,r).
\end{align}
As illustrated in Fig.~\ref{fig:K-vs-L-q},
the curves for the channel length dependence
of the the asymptotic secret fraction
lie within the region
bounded by $K^{(\ind{DH})}_{\ind{UB}}(l)$
and
$K^{(\ind{DH})}_{\ind{LB}}(l)$.
It is seen that
the size of this region increases with the imbalance ratio, $q$.

% \textcolor{blue}{Fig.~\ref{fig:K-vs-L-q} shows the dependence of asymptotic secret fraction on
%   channel length, as well as upper and lower bounds that arise from squeezing dependence of joint
%   entropy. In this figure, quadrature variances are fixed, $\sigma_x^{(i)}=1.01$, and the asymmetry
%   is changed by varying the imbalance ratio $q$~\eqref{eq:q-param}. For low asymmetries, the effect
%   in negligible (less than $0.5$ km difference), as seen from Fig~\ref{subfig:K-vs-L-q-a}. As the
%   asymmetry increases, corresponding to the increase of the length of the $r$-interval, range of
%   possible values of reach increases as well, as illustrated by Fig~\ref{subfig:K-vs-L-q-b}. In this
%   case, the dependence of joint entropy on squeezing parameter~\eqref{eq:r1-r2} becomes
%   non-negligible.}

Since in the scenario under consideration,
Eve controls the measurement noise channel,
the freedom to choose the value of $r$
might be used to maximize the information available to
the eavesdropper. This strategy is equivalent to
maximizing the entropy $S_{\ind{AB}}^{(\ind{DH})}$
that leads to the coherent-state representation
of double homodyne POVM only if
$q=1$ and $\sigma_x^{(1)}=\sigma_x^{(2)}$.
In other cases, the optimal value of $r$ will differ  from zero and corresponds to
the squeezed-state representation~\eqref{eq:Pi_G-gen-x1x2}.

% If
% input beamplitter is unbalanced (red curve in
% Fig.~\ref{fig:Entropy}b), interval of possible $r$, as well as optimal
% squeezing parameterthat maximazis Holevo information, shift.

% Moreover,
% if disbalance of input beam splitter $\left|1-q\right|$ is sufficiently
% high, choice of $r=0$ becomes impossible as it does not lay in the
% interval~\eqref{eq:r1-r2}. Substituing $r\not\in[r_2,r_1]$ results in
% one of eigenvalues being less than $1$, which is
% unphysical. Therefore, the more general representation given by
% Eq.~\eqref{eq:Pi_G-gen-x1x2} must be employed.
% %unequal sigma_x^i

% In asymmetrical case (blue curves) if input beam splitter is balanced,
% while using $r=0$, i.e. Eq.~\eqref{eq:povm-hetero}, is possible, joint
% entropy is maximized by $r\neq0$. If input beam splitter is unbalanced,
% similarly to red curve, dependencies of $r$ shift in accordance to
% disbalance ratio.

\begin{figure*}[!htb]
    \centering
    \begin{subfigure}[c]{.3\linewidth}
\includegraphics[width=\linewidth, ]{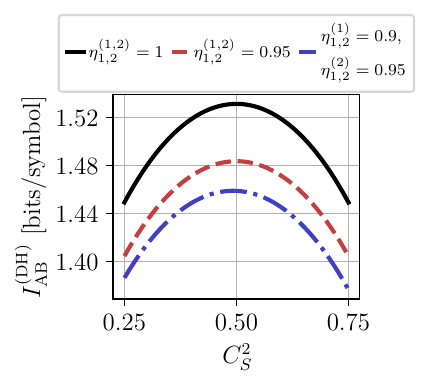}
\caption[]{}%{$I_{\ind{AB}}$}
\label{subfig:dhom-a}
        \end{subfigure}
\hfill
        \begin{subfigure}[c]{.3\linewidth}
 \includegraphics[width=\linewidth, ]{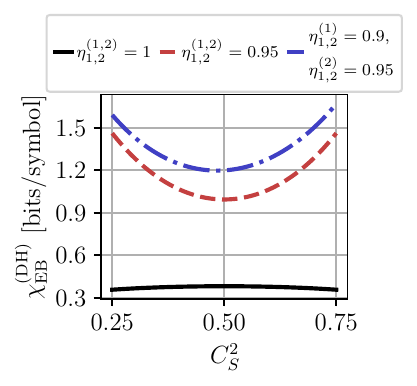}
 \caption[]{}%{$\chi_{\ind{EB}}$}
 \label{subfig:dhom-b}
\end{subfigure}
\hfill
    \begin{subfigure}[c]{.3\linewidth}
\includegraphics[width=\linewidth, ]{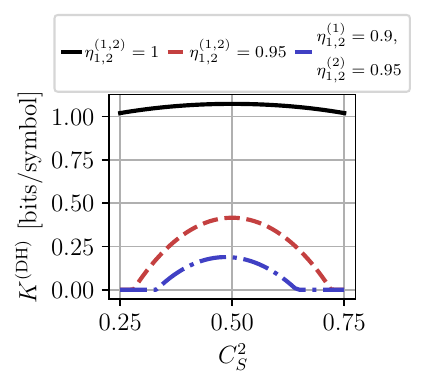}
\caption[]{}%{$R$}
\label{subfig:dhom-c}
        \end{subfigure}
        \caption{Informational quantities computed for the case of double-homodyne receiver (see Fig.~\ref{fig:double-homodyne}). 
          (a)~Mutual information, $I_{\ind{AB}}^{(\ind{DH})}$, (b)~Holevo information,
          $\chi_{\ind{EB}}^{(\ind{DH})}$ and (c)~asymptotic secret fraction, $K^{(\ind{DH})}$, as a
          function of the signal beam splitter transmission, $C_S^2$, computed at different values of the detector
          efficiencies.  All other beam splitters are assumed to be balanced and efficiencies not
          specified in the legend are taken to be unity.  The parameters are: $V=5$; $T=0.95$;
          $\xi=10^{-2}$; and $\beta=0.95$.  }
\label{fig:double-homodyne-all}
\end{figure*}

\begin{figure}
    \centering
\includegraphics[width=\linewidth]{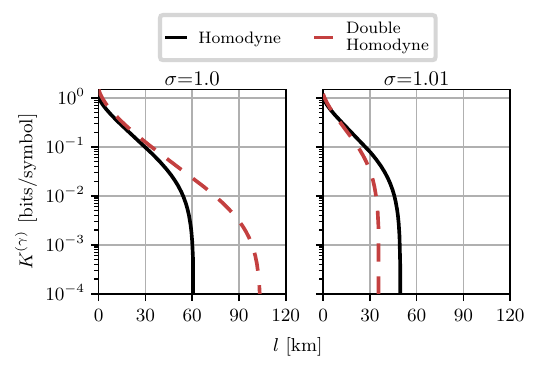}
\caption{Asymptotic secret fractions for homodyne and double-homodyne
  receivers, $K^{(\ind{H})}$ and $K^{(\ind{DH})}$, respectively, as a function of the
  channel length, $l$,
  computed at $\sigma_x=
  \sigma_1=\sigma_2\in\{1.0,1.01\}$. The losses are assumed to be $20$ dB per $100$ km.  The
  parameters are: $V=5$; $\beta=0.95$; $\xi=0.5\times10^{-3}$.
  For double homodyne detection, the squeezing parameter is taken to be zero.}
      \label{fig:K-comparison}
      \end{figure}

The graphs presented in Fig.~\ref{fig:double-homodyne-all}
show the mutual information, the Holevo information
and the secret key fraction
computed as a function
of the input (signal) beam splitter transmittance
of the double homodyne scheme (see Fig.~\ref{fig:double-homodyne})
In these calculations,
the value of the squeezing parameter is taken to be optimal
so that the Holevo information is maximized at
$r=r_{\ind{opt}}=\argmax_r\chi_{\ind{EB}}^{(\ind{DH})}(r)$. 

Qualitatively,
the curves plotted in Fig.~\ref{fig:double-homodyne-all}
reveal a behavior similar to
that evaluated for homodyne detection (see Fig.~\ref{fig:homodyne-all}). 
A comparison between Fig.~\ref{subfig:hom-a}
and Fig.~\ref{subfig:dhom-a} shows that, as expected,
the mutual information for double homodyne detection 
generally exceeds the information for homodyne detection,
$I_{\ind{AB}}^{(\ind{DH})}>I_{\ind{AB}}^{(\ind{H})}$.
For noiseless detection with $\sigma_x^{(1)}=\sigma_x^{(2)}=1$
(black solid curves in Figs.~\ref{fig:double-homodyne-all} and~\ref{fig:homodyne-all}),
the resulting double-homodyne asymptotic secret fraction plotted in Fig.~\ref{subfig:dhom-c}
is noticeably higher than  the value computed for
homodyne detection shown in Fig.~\ref{subfig:hom-c},
$K^{(\ind{DH})}>K^{(\ind{H})}$.
In this case, 
an interesting feature illustrated
by the black curve for the Holevo information
(see Fig.~\ref{subfig:dhom-b})
is that the imbalance of the input (signal) beam splitter
results in decreased Holevo information.
The reason is that, in the limit of ideal detection,
the excess noise variances, $\sigma_{N}^{(1)}$ and
$\sigma_{N}^{(2)}$, are equal to zero
and the entropy $S_{\ind{AB}}^{(\ind{DH})}$ is independent of $C_S^2$,
whereas the conditional entropy $S_{\ind{A|B}}^{(\ind{DH})}$,
being an even function of $1/2-C_S^2$,
grows with imbalance.

When measurements are noisy,
the dashed and dash-dotted curves
plotted in Figs.~\ref{subfig:hom-b} and~\ref{subfig:dhom-b}
indicate a noise induced increase
of the Holevo information
so that its value
for double homodyne detection
appears to be
significantly larger than the homodyne Holevo information.
As a consequence
it turned out that
the double-homodyne secret fraction $K^{(\ind{DH})}$
is lower than $K^{(\ind{H})}$
(see Figs.~\ref{subfig:hom-c} and~\ref{subfig:dhom-c}).

Figure~\ref{fig:K-comparison} presents
the secret key fraction-vs-distance curves for both the homodyne and
double homodyne measurements
computed at $\sigma_x=\sigma_{1}=\sigma_{2}\equiv\sigma$.
Clearly,
in the case of ideal measurements with
$\sigma=1$,
the largest asymptotic secret fraction corresponds to double homodyne detection.
When the measurements are
        noisy with $\sigma>1$, the double homodyne scheme having higher asymptotic secret key fraction at
        short distances is outperformed by the homodyne detection at longer distances.

%For further calculations, we will optimize $r$ to maximize $\chi_{\ind{EB}}$.

% The effects of detection asymmetry on mutual information, Holevo
% information, and the asymptotic secret fraction are illustrated in
%  for double homodyne detection.

% From Fig.~\ref{fig:homodyne-all}, it is evident that while asymmetry
% in homodyne detection generally degrades performance, the combination
% of an unbalanced beam splitter and unequal detector efficiencies can be
% beneficial, resulting in maxima of the mutual information and secret
% fraction with given parameters. Correspondingly, the minima of the
% Holevo information exhibit similar
% behavior. % Notably, these functions are symmetric and even with shifting local mininimum (symmetry point).
% Our numerical results for the homodyne case are qualitatively
% consistent with Ref.~\cite{ruiz2023effects}, namely the dependence of
% the asymptotic secret fraction on the beam splitter transmittance
% deviation and balanced detector deviation.

% The discrepancy in our
% Holevo
% information calculations leads to results for the asymptotic secret
% fraction dependence under double homodyne detection that differ from
% those in }, although both works reach the same
% conclusion regarding its 

% \cite{Lupo:prxq:2022}

%%%%%%%%%%%%%%%%%%%%
\section{Discussion and conclusion}
\label{sec:conclusion}
%%%%%%%%%%%%%%%%%%%%%%%%%%%%

In this paper, we have studied the effects of asymmetry introduced by unbalanced beam splitters and
different efficiencies of the photodetectors in photocount statistics of homodyne and double
homodyne detection.
By using the Gaussian approximation
derived by approximating the Poisson distributions with the probability density
functions of normally distributed random variables,
we have computed the $Q$ symbols of
the POVMs of noisy homodyne and double homodyne (heterodyne)
measurements.
(The applicability range of the Gaussian
approximation is discussed in Appendix~\ref{sec:appendix_numerical}.)
Similarly to Gaussian states, these POVMs are generally characterized
by the mean quadratures and the covariance matrix.

For homodyne measurements,
we have used the $Q$ symbol~\eqref{eq:Pgood-w-def}
to deduce the homodyne covariance matrix~\eqref{eq:Sigma-H-m}.
The operator representation of the noisy homodyne POVM~\eqref{eq:homodyne-povm}
is derived by applying the additive
noise (classical mixing) channel~\eqref{eq:Phi_N}
with the excess noise covariance matrix~\eqref{eq:Sigma-N-H}
to the POVM of
the noiseless (ideal) measurements~\eqref{eq:POVM-P0}
described by the projectors onto the quadrature eigenstates~\eqref{eq:ket-x-phi}.
This representation arises as a natural result of our general approach,
which is outlined in Sec.~\ref{subsec:hom-POVM}
and is based on the well-known fact that
a POVM of a noisy Gaussian measurement
can be obtained as
a POVM of a projective Gaussian measurement
transformed under the action of the dual of a suitably chosen Gaussian channel~\cite{Serafini:bk:2023}. 

When this approach is applied to the case of double
homodyne measurements
(the asymmetric eight-port  double homodyne scheme is illustrated in Fig.~\ref{fig:double-homodyne})
with the $Q$ symbol~\eqref{eq:P_G-x1x2}
and the covariance matrix~\eqref{eq:Sigma_m-DH},
we have shown that
the class of noiseless measurements
should be extended so as to include
the squeezed-state projective measurements
(these states are given by Eq.~\eqref{eq:squeezed-def})
with the squeezing parameter
ranged between the endpoints of the interval~\eqref{eq:r1-r2}.
It was found that
the squeezed-state operator representation  of the double-homodyne POVM~\eqref{eq:Pi_G-gen-x1x2}
emerges from such POVMs of noiseless squeezed-state measurements
under the action of
the additive noise channel
with the excess noise covariance matrix~\eqref{eq:Sigma-N-DH}.
So, the asymmetry induced anisotropy of
the distributions in the photocount difference plane illustrated in
Fig.~\ref{fig:dh-statistics}
leads to the ambiguity of the operator representation of the double-homodyne POVM
that appears to depend on the squeezing parameter.

We have demonstrated that
this inherent non-uniqueness of the POVM representation
becomes relevant
when analyzing the asymptotic security
of the Gaussian-modulated CV-QKD protocol
in the untrusted-noise scenario.
Specifically, we have studied
how the asymmetry of the noisy measurements
impacts the performance of CV-QKD system in
the scenario where the measurement noise is accessible to Eve.

In this scenario, the upper bound for the Holevo information is estimated by assuming that the
covariance matrix of the state shared by Alice and Bob is additionally transmitted through the
channel describing the noise of Bob's measurements.
Since this state is purified by Eve, the Eve's (joint)
entropy explicitly depends on the parameters of the measurement noise channel.
In our case, this
channel is chosen to be the additive noise channel~\eqref{eq:Phi_N} with
the excess noise covariance matrix given by Eqs.~\eqref{eq:Sigma-N-H}
and~\eqref{eq:Sigma-N-DH} for homodyne and double homodyne measurements,
respectively.
It was found that, for both types of measurements,
the mutual information, the Holevo information and the asymptotic secret fraction
are sensitive to asymmetry effects (see Figs.~\ref{fig:homodyne-all}
and~\ref{fig:double-homodyne-all}).
The imperfections also cause a degraded performance
of the protocol leading to a significant reduction in the maximum channel length
(see Fig.~\ref{fig:K-homodyne} and Fig.~\ref{fig:K-DH-r}).

A distinguishing feature of the double homodyne detection is that we have to deal with the squeezing
dependent joint (Eve's) entropy (see Fig.~\ref{fig:Entropy}).
In the symmetric case,
it turned out that
the standard coherent-state representation
with vanishing squeezing parameter, $r=0$,
maximizes both the Eve's entropy and the Holevo information.
In the presence of asymmetry, the coherent-state point
can be outside the interval where the entropy is well-defined.
In this case,
the results presented in Fig.~\ref{fig:K-vs-L-q}
show that
the lower and upper bounds for the secret fraction
(see Eq.~\eqref{eq:K_LU})
are sensitive to the imbalance ratio~\eqref{eq:q-param}.

% Since Eve controls the channel, the
% latter should be optimized.  For the curves presented in Fig.~\ref{fig:double-homodyne-all}, the
% optimal values of the squeezing parameter are found by maximization of the Holevo information.

% \textcolor{blue}{While in a typical setup standard description that uses $r=0$ is accurate (see
%   Fig.~\ref{fig:K-DH-r}), when asymmetry is significant, the effect becomes non-negligible (see
%   Fig.~\ref{fig:K-vs-L-q}).}

% \textcolor{blue}{Comparing symmetrical schemes in the untrusted noise scenario revealed that
%   imperfect double homodyne performs worse than homodyne (see Fig.~\ref{fig:K-comparison}). Although
%   mutual information is higher with double homodyne detection, so is the Holevo information. This
%   result aligns with the analysis of perfect (in the sense of this work) detection schemes in
%   Ref.~\cite{laudenbach2018continuous}, which states that double homodyne is more sensitive to
%   excess noise.}

In this study,
for illustrative purposes, we have restricted our attention to 
the case of the untrusted-noise scenario.
The opposite case of
the trusted-noise (device)
scenario~\cite{Lodewyck:pra:2007,Usenko:entropy:2016,Laudenbach:aqt:2019}
is beyond the scope of this paper
as it requires
an additional and a more sophisticated analysis, which is the subject of a separate publication.

In conclusion,
we put our results in the general context of CV-QKD
  security~\cite{Zhang:apr:2024}.
  The asymmetry effects described in the paper, such as
deviations from the ideal 50:50 beam splitter ratio and mismatched detector efficiencies,
introduce vulnerabilities into CV-QKD systems.  These flaws  allow quantum hackers not only to
compromise security, but also to degrade system performance~\cite{Ruiz:heliyon:2023}.  These
vulnerabilities can be exploited by adversaries through attack strategies such as the wavelength
attack~\cite{huang2012wavelength, huang2014quantum} and the homodyne detector blinding~\cite{qin2018homodyne} and saturation~\cite{qin2016quantum}. The wavelength attack leverages the
wavelength-dependent coupling ratio of fiber beam splitters and can be countered by using proper
spectral filtering. Blinding and saturation attacks exploit the saturation behavior of homodyne
detectors, and their effectiveness is amplified by receiver imbalance. If the splitting ratio
deviates from  the ideal one, an injected bright pulse more easily displaces the detector output,
facilitating saturation and biasing excess noise estimation.

%%%%%%%%%%%%%%%%%%%%%%%%%%%%%%
\section*{Acknowledgements}
The work was supported by the Russian Science Foundation (project No. 24-11-00398).
%%%%%%%%%%%%%%%%%%%%%%%%%%%%%%%%%

\appendix
\section{Statistical distance between Gaussian approximation and Skellam distribution}
\label{sec:appendix_numerical}

\begin{figure*}[!htb]
  \centering
  \begin{subfigure}{0.49\textwidth}
    \includegraphics[width=.9\linewidth]{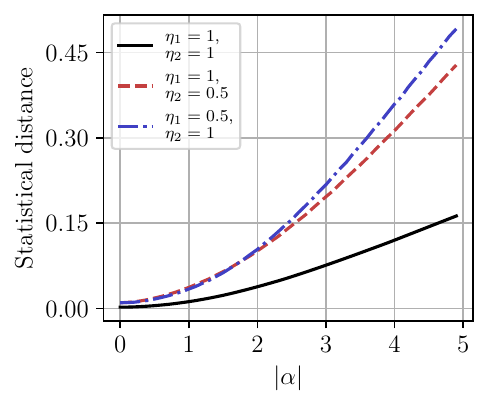}
    \caption{$\left|\alpha_L\right|=5$}
    \label{fig:amp_05}
\end{subfigure}
\hfill
  \begin{subfigure}{0.49\textwidth}
    \includegraphics[width=.9\linewidth]{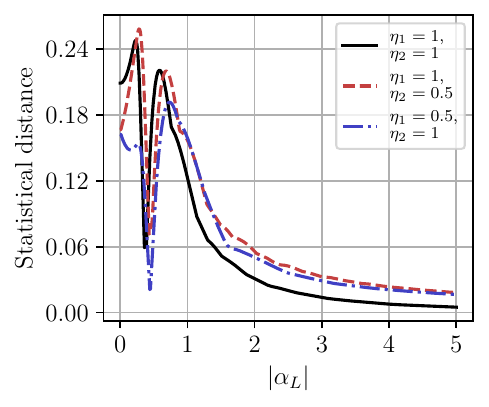}
    \caption{$|\alpha|=0.5$}
    \label{fig:amp_01}
\end{subfigure}
\caption{
  Statistical distance $D_P=\mathtt{D}(\prob,\prob_G)$
  as a function of (a)~signal amplitude and (b)~LO amplitude at different
          detector efficiencies.
        }
  \label{fig:amplitude}
\end{figure*}

\begin{figure*}
    \centering
    \begin{subfigure}[b]{.45\linewidth}
\includegraphics[width=\linewidth]{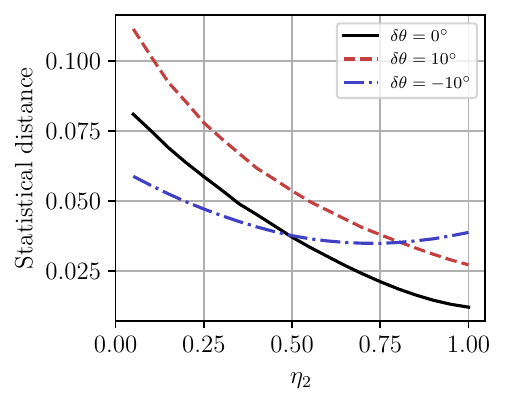}
\caption[]{$\eta_1=1$}
\label{fig:eta1-1}
        \end{subfigure}
        \hfill
\begin{subfigure}[b]{.45\linewidth}
 \includegraphics[width=.97\linewidth]{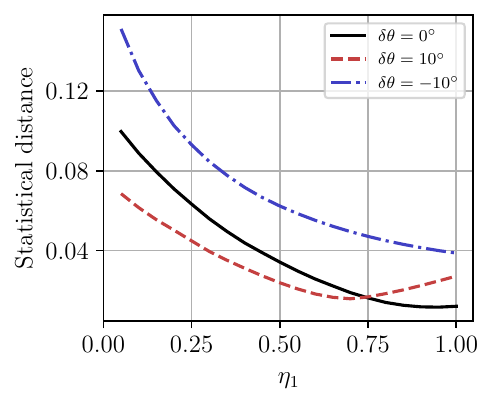}
 \caption[]{$\eta_2=1$}
 \label{fig:eta2-1}
\end{subfigure}
\caption{Statistical distance $D_P=\mathtt{D}(\prob,\prob_G)$
  as a function of  photodetector efficiency
  (a)~$\eta_2$ at $\eta_1=1$ and
  (b)~$\eta_1$ at $\eta_2=1$
  for different values of the beam splitter imbalance angle
  $\delta\theta$ (see eq.~\eqref{eq:delta-theta}).
  The amplitudes are
  $|\alpha|=1$ and $\left|\alpha_L\right|=5$.
}
\label{fig:delta-eta}
\end{figure*}

In this Appendix, we study the perfomance of Gaussian approximation for the statistics of photon count difference~\eqref{eq:Gaussian}:
\begin{align}
  &
  \label{eq:App_Gaussian}
    \prob_G(\mu)=G(\mu-\mu_G;\sigma_G),
    \\
  &
    \label{eq:App_sigma_G}
    \sigma_G=\eta_1|\alpha_1|^2+\eta_2|\alpha_2|^2
    \approx
    (\eta_1S^2+\eta_2C^2) \left|\alpha_L\right|^2,
  \\
  &
    \label{eq:App_mu_G}
    \mu_G=\eta_1|\alpha_1|^2-\eta_2|\alpha_2|^2
    \approx
    (\eta_1S^2-\eta_2C^2) \left|\alpha_L\right|^2
    \notag
  \\
  &
    +CS (\eta_1+\eta_2) \left|\alpha_L\right| \avr{\hat{x}_\phi}
\end{align}
by comparing it with the exact statistics of difference events
governed by the Skellam distribution~\eqref{eq:accurate}:
\begin{align}
  \label{eq:App_accurate}
  \prob(\mu)&=e^{-\eta_1|\alpha_1|^2}e^{-\eta_2|\alpha_2|^2}
    \Biggl(\frac{\eta_1|\alpha_1|^2}{\eta_2|\alpha_2|^2}\Biggr)^{\mu/2}
    \notag
  \\
  &
    \times
I_{\mu}\bigl(2\sqrt{\eta_1\eta_2|\alpha_1|^2|\alpha_2|^2}\bigr),
\end{align}
both repeated here for ease. We limit our numerical results to the coherent signal state only, except in cases where the curves show sufficiently distinct differences.

We evaluate the statistical distance between the probability distributions using the total variational distance
that can be computed as half of the $L^1$ distance
\begin{align}
  &
    \label{eq:stat-dist}
    D_P\equiv\mathtt{D}\left(\prob,\prob_{G}\right)\equiv
    \frac{1}{2}
    \sum_{\mu=-\infty}^{\infty}
    \left\lvert\prob(\mu) -\prob_G(\mu)\right\rvert.
  \end{align}
  Note that, according to Eq.~\eqref{eq:stat-dist},
  $\mu$ takes integer values and
  we evaluate the distance between the probability mass fuctions,
  whereas the normalization condition for
the Gaussian function~\eqref{eq:App_Gaussian}
\begin{align}
    \label{eq:norm-cont}
  \int_{-\infty}^{\infty}\mathop{\dd\mu}
    \prob_G(\mu)
    =1
\end{align}
implies applicability of the continuum limit.
For integer $\mu$,
the integral
on the left hand side of Eq.~\eqref{eq:norm-cont}
should be replaced with a sum
and we have the relation
\begin{align}
  \label{eq:discr-norm}
  \sum_{\mu=-\infty}^{\infty} \prob_G(\mu)=\vartheta_3(\pi\mu_G,e^{-2\pi^2\sigma_G})\equiv N_G
\end{align}
where $\vartheta_3$ is the Jacobi elliptic theta function~\cite{NIST:hndbk:2010}.

In the applicability region of
the continuum limit, the normalization constant $N_G$
is close to unity. The numerical analysis shows that
$|N_G-1|\le 10^{-4}$ at $ 2\sigma_{G}\ge 1$.
The latter gives the condition for the LO amplitude
\begin{equation}
    \left|\alpha_L\right|\geq \frac{1}{\sqrt{2(\eta_1S^2+\eta_2C^2)}}\equiv\alpha_N
    \label{eq:renorm}
  \end{equation}
   which ensures both applicability of the continuum limit
  and proper normalization of the Gaussian approximation.
  In our calculations,
the probability $\prob_G$ will be numerically corrected by introducing
the factor $N_G^{-1}$
provided that $\left|\alpha_L\right|$ is below the "renormalization point" $\alpha_N$.

  The curves presented in Fig.~\ref{fig:amplitude}
  illustrate how the accuracy of the Gaussian approximation
  is affected by the signal and LO amplitudes, $|\alpha|$ and $\left|\alpha_L\right|$.
  More specifically, in
Fig.~\ref{fig:amp_05} (Fig.~\ref{fig:amp_01}),
  the statistical distance is numerically evaluated as a function
  of the amplitude $|\alpha|$ ($\left|\alpha_L\right|$)
at different values of the photodetectors efficiencies
provided that the value of
the other amplitude $\left|\alpha_L\right|$ ($|\alpha|$) is fixed.

\begin{figure*}
    \centering
    \begin{subfigure}[b]{.45\textwidth}
\includegraphics[width=\linewidth]{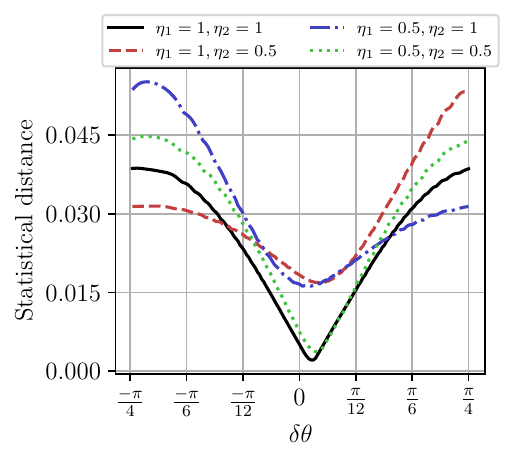}
\caption[]{$\ket{\psi}=\ket{\alpha},\:\alpha=0.5$}
\label{fig:dth_alp}
        \end{subfigure}
        \hfill
        \begin{subfigure}[b]{.45\textwidth}
 \includegraphics[width=.97\linewidth]{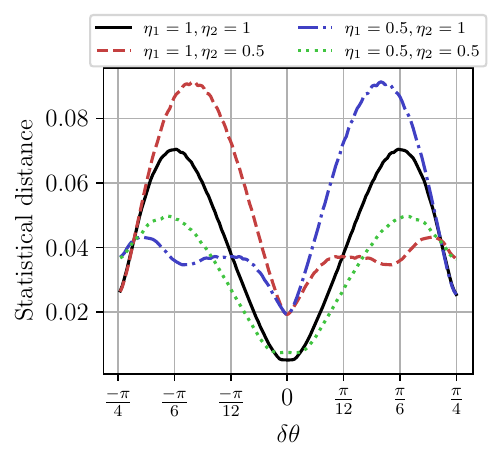}
\caption[]{$\ket{\psi}=\ket{n},\:n=1$}
\label{fig:dth_fock}
\end{subfigure}
\caption{
  Statistical distance $D_P=\mathtt{D}(\prob,\prob_G)$
  as a function of the beam splitter imbalance angle, $\delta\theta$,
for the signal mode prepared in (a)~the coherent state and
  in (b)~the single photon Fock state at different efficiencies  with $\left|\alpha_L\right|=5$.
    }
    \label{fig:delta-theta}
\end{figure*}

Referring to Fig.~\ref{fig:amp_05},
the curves behave as expected:
given the LO amplitude $\left|\alpha_L\right|$,
the distance monotonically increases with $|\alpha|$.
It is shown that,
at $\left|\alpha_L\right|=5$ and $|\alpha|>1$,
the perfectly symmetric homodyne presents the case
with minimal distance, $D_P$, while in the presence of asymmetry,
the curves exhibit a rapid growth and
the quality of  the Gaussian approximation
rapidly degrades to the point, where $D_P>0.1$,
so it is not useful for its intended purpose.

When it comes to dependencies of the statistical distance on
the LO oscillator amplitude
computed at fixed value of $|\alpha|$,
the above results suggest that the smaller the amplitude $|\alpha|$
the better the accuracy of the Gaussian approximation.
We can also expect the distance
will be small provided that $\left|\alpha_L\right|$ is large
and the strong-LO approximation is applicable.

Referring to Fig.~\ref{fig:amp_01},
the curves evaluated at $|\alpha|=0.5$
display a non-monotonic behavior with two local maxima
in the weak LO range where $\left|\alpha_L\right|<1$.
By contrast, after the second maximum at $\left|\alpha_L\right|>1$,
the distance falls with the LO amplitude
and it drops below $0.05$ at $\left|\alpha_L\right|>2$.

Note, that, when $|\alpha|<0.1$
and the signal mode state is close to the vacuum state,
the two local maxima of $D_P$ can be estimated to be slightly above $0.08$ and $0.1$,
respectively. So, in this case, the distribution~\eqref{eq:Gaussian}
might be regarded as a reasonable approximation even in the
weak LO range where
the probability $\prob_G$ approaches the close neighborhood of
the singular limit,
$\lim_{\left|\alpha_L\right|\to 0}\prob_G(\mu)=\delta(\mu)$.

From Fig.~\ref{fig:amp_01},
it can also be seen that
the distance vs LO amplitude dependence
that can be used as a tool to characterize
the applicability region of the strong-LO approximation
is nearly insensitive to asymmetry.
In other words, the latter does not produce noticeable effects on
the accuracy of the approximation.

The parameters describing
the photodetection asymmetry are
the efficiencies $\eta_1$ and $\eta_2$.
The curves plotted in Fig.~\ref{fig:amplitude}
are computed at different values of the efficiencies.

To quantify deviation of the beam splitter
transmission and reflection amplitudes
from the balanced $50:50$ values
$t=\cos\theta=r=\sin\theta=1/\sqrt{2}$ at
the angle $\theta=\pi/4$,
we introduce the beam splitter \textit{imbalance angle} given by
\begin{equation}
  \label{eq:delta-theta}
        \delta\theta\equiv\frac{\pi}{4}-\theta.
      \end{equation}

      In Figure~\ref{fig:delta-eta}
      we plot the distance against
      the efficiency of the photodetector assuming that the
      other photodetector is perfect.
      The curves are evaluated at different values of the
      imbalance angle~\eqref{eq:delta-theta}.

In Fig.~\ref{fig:delta-theta} 
the statistical distance vs imbalance angle
curves are presented for coherent and one-photon signal states. 
These curves illustrate how the beam splitter imbalance and
the photodetector efficiencies influence the accuracy of the
Gaussian approximation.
The distance is shown to be minimal in the vicinity of
the balanced beam splitter point with
a vanishing imbalance angle, $\delta\theta=0$.
The efficiency dependence of
the distance
is shown to decrease monotonically at $\delta\theta=0$.
In contrast, for unbalanced beam splitter, this dependence can
reveal non-monotonic behavior.

% %%%%%%%%%%%%%%%%%%%%%%%%%%%%%%
\section{Gaussian approximation from Skellam distribution}
\label{sec:appendix_comparison}

The derivation procedure for the Gaussian approximation
outlined in Sec.~\ref{sec:homodyne} transforms
the photocount difference probability~\eqref{eq:poisson}
into the form of a convolution of the normal distributions
by approximating the Poisson distributions.
In this Appendix,
we discuss an alternative method
where the starting point is the Skellam distribution~\eqref{eq:accurate}.
For convenience, we shall reproduce the expression for
this distribution here:
\begin{align}
\label{eq:Skellam}   
  P(\mu)&=e^{-\eta_1|\alpha_1|^2}e^{-\eta_2|\alpha_2|^2}
  \left(\frac{\eta_1|\alpha_1|^2}{\eta_2|\alpha_2|^2}\right)^{\frac{\mu}{2}}
  \notag
  \\
  &
  \times
I_{\mu}\left(2\sqrt{\eta_1\eta_2|\alpha_1|^2|\alpha_2|^2}\right),
\end{align}
where $I_k(z)$ is the modified Bessel function of the first kind
and the amplitudes $|\alpha_{1,2}|$ are given by Eq.~\eqref{eq:amplitudes}.

The method under consideration
(see, e.g. the textbook~\cite{Vogel:bk:2006})
assumes that,
in the strong LO limit, the argument of the modified Bessel function is large
and $I_\mu(z)$ can be
approximated using its asymptotic expansion
taken in the Gaussian form:
\begin{align}
  \label{eq:asymp-A}
  I_\mu(z)\approx\frac{1}{\sqrt{2\pi z}}\exp\left[z-\frac{\mu^2}{2 z}\right].
\end{align}
This formula can be deduced by
performing a saddle-point analysis
for the integral representation of
the Bessel functions~\cite{freyberger1993photon}.

Heuristically, it can also be obtained from
from the lowest order asymptotic expansion for
the Bessel function~\cite{NIST:hndbk:2010}:
$I_{\mu}(z)\approx e^z (1-(4\mu^2-1)/(8z))/\sqrt{2\pi z}$
assuming that, for small values of $x$,
$1-x$ can be replaced with $e^{-x}$
(the factors independent of $\mu$ are not essential because they can be incorporated
into the normalization factor of the Gaussian approximation).

\begin{figure}[ht!]
    \centering
    \includegraphics[width=.75\linewidth]{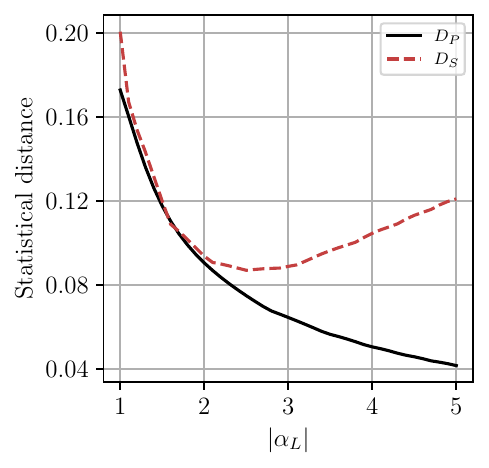}
    \caption{Distances, $D_P=D(\prob,\prob_G)$ and $D_S=D(P,{\prob}_G^{(s)})$, 
      between the Skellam distribution, $\prob$, and
      two Gaussian approximations, $\prob_G$ (see Eq.~\eqref{eq:Gaussian})
      and  ${\prob}_G^{(s)}$ (see Eq.~\eqref{eq:tPG-mu}), as a function of $\left|\alpha_L\right|$
      the beam splitter imbalance angle  at $\delta\theta=15^{\circ}$
$\alpha=1$, and $\eta_1=\eta_2=1$.
      % Note the significant difference (in first decimal places) between the curves, which means that $P_B$
      % becomes incorrect for higher asymmetry
    }
    \label{fig:PGPGs-aL}
\end{figure}

Assuming that $CS\ne 0$ and $\left|\alpha_L\right|$ is sufficiently large,
we can use the approximate relations
\begin{align}
  &
    \label{eq:z1-A}
    z=2\sqrt{\eta_1\eta_2}|\alpha_1||\alpha_2|
    \approx
    2CS\sqrt{\eta_1\eta_2}\left|\alpha_L\right|^2,
  \\
  &
    \label{eq:z2-A}
    \ln \left(\frac{{\eta_1}|\alpha_1|^2}{{\eta_2}|\alpha_2|^2}\right)^{\frac{\mu}{2}}\approx
    \frac{\mu}{2}\left(
    \ln\frac{\eta_1S^2}{\eta_2C^2}+\frac{\langle\hat{x}_\phi\rangle}{CS\left|\alpha_L\right|}\right)
    \end{align}
    to obtain the Gaussian approximation for the Skellam distribution~\eqref{eq:Skellam}
    given by
\begin{align}
  &
  \label{eq:tPG-mu}
        {\prob}_G^{(s)}(\mu)=G(\mu-\tilde{\mu}_G;\tilde{\sigma}_G),
  \\
  &
  \label{eq:tmu_G}
  \tilde{\mu}_G=\sqrt{\eta_1\eta_2}\Bigl[
  CS \left|\alpha_L\right|^2\ln\frac{\eta_1S^2}{\eta_2C^2}+
  \left|\alpha_L\right|\avr{\hat{x}_{\phi}}
  \Bigr],
  \\
  &
      \label{eq:tsgm_G}
    \tilde{\sigma}_G=
    2CS\sqrt{\eta_1\eta_2}\left|\alpha_L\right|^2.
\end{align}

Similar to Eq.~\eqref{eq:Pgood-w-def},
we cast the probabilty~\eqref{eq:tPG-mu}
into the following quadrature form: 
\begin{align}
  &
    \label{eq:tPG-tx}
{\prob}_G^{(s)}(\tilde{x})=\frac{1}{\sqrt{2\pi\tilde{\sigma}_G}}
    \exp \biggl\{-\frac{(\tilde{x}-\avr{\hat{x}_\phi})^2}{2\tilde{\sigma}_x}\biggr\},
  \\
  &
    \label{eq:tld-x}
    \tilde{x}=\frac{\mu}{{\sqrt{\eta_1\eta_2}\left|\alpha_L\right|}}-
    CS\left|\alpha_L\right|\ln\frac{\eta_1S^2}{\eta_2C^2},
    \\
  &
    \label{eq:tsgm_x}
    \tilde{\sigma}_x=\frac{2CS}{\sqrt{\eta_1\eta_2}},
\end{align}
so that we may follow the line of reasoning
presented in Sec.~\ref{sec:homodyne} to deduce the POVM
\begin{align}
\label{eq:tpovm}    
    \hat{\Pi}_G^{(s)}&=\frac{1}{\sqrt{\eta_1\eta_2}\left|\alpha_L\right|}
    \notag
  \\
  &
    \times
    \int \dd x' G(x-x'; \tilde{\sigma}_N)|x',\phi\rangle\langle x',\phi|
\end{align}
with the noise variance
\begin{equation}
  \label{eq:tsigm-N}
  \tilde{\sigma}_N=\tilde{\sigma}_x-1,
  \quad
  0 \le \tilde{\sigma}_x\le \tilde{\sigma}_x^{(\max)}=1/\sqrt{\eta_1\eta_2}.
\end{equation}

\begin{figure}[ht!]
    \centering
    \includegraphics[width=.75\linewidth]{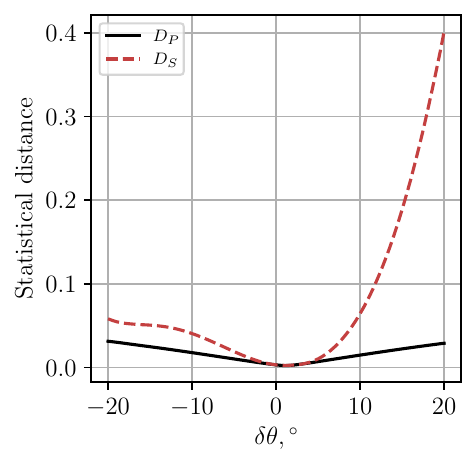}
    \caption{Distances, $D_P=D(\prob,\prob_G)$ and $D_S=D(\prob,\prob_G^{(s)})$, 
      between the Skellam distribution, $P$, and
      two Gaussian approximations, $\prob_G$ (see Eq.~\eqref{eq:Gaussian})
      and  ${\prob}_G^{(s)}$ (see Eq.~\eqref{eq:tPG-mu}), as a function of
      the beam splitter imbalance angle $\delta\theta$ at
$\alpha=1$, $\alpha_L=10$ and
      $\eta_1=\eta_2=1$.
      % Note the significant difference (in first decimal places) between the curves, which means that $P_B$
      % becomes incorrect for higher asymmetry
    }
    \label{fig:PGPGs-dth}
\end{figure}

Note that,
in the symmetric
case with $C=S$ and $\eta_1=\eta_2$,
the Gaussian distributions given by
Eqs.~\eqref{eq:Gaussian} and~\eqref{eq:tPG-mu} are equivalent.
This is no longer the case in the presence of asymmetry effects.

Figure~\ref{fig:PGPGs-aL} demonstrates
that, at $\delta\theta\ne 0$,
by contrast to the distance between $P$ and $\prob_G$,
$D_P=\mathtt{D}(\prob,\prob_G)$ which is a monotonically decreasing function of
$\left|\alpha_L\right|$,
the distance
between the Skellam distribution and
the approximate distribution~\eqref{eq:tPG-mu},
$D_S=\mathtt{D}(\prob,\prob_G^{(s)})$,
reveals non-monotonic behaviour and
increases with $\left|\alpha_L\right|$ at sufficiently large LO amplitudes.
Referring to Fig.~\ref{fig:PGPGs-dth},
imbalance of the beam splitter
has strong detrimental effect on
the accuracy of the approximation~\eqref{eq:tPG-mu}.

What is more important is
that, by contrast the noise excess variance~\eqref{eq:sgm_N}
which is always positive,
the variance~\eqref{eq:tsigm-N} becomes negative
when $2CS\le \sqrt{\eta_1\eta_2}$.
The latter breaks applicability of Eq.~\eqref{eq:tpovm}
giving an ill-posed POVM.

%%%%%%%%%%%%%%%%%%%%%%%%%%%%%%%%%%%%%%%%%%%
\section{Mutual information and input state
  of entanglement-based representation}
\label{sec:MI-and-EBtoPm}
%%%%%%%%%%%%%%%%%%%%%%%%%%%%%%%%%%%%%%

In this appendix,
we begin with derivation of
the expressions
for the mutual information
between Alice and Bob
in the Gaussian-modulated coherent-state
CV-QKD protocol discussed in Sec.~\ref{sec:protocol}.
These expressions are given by Eqs.~\eqref{eq:I_H_res}
and~\eqref{eq:I_DH_res}.

The important point greatly simplifying calculations
is that both the channel~\eqref{eq:AB-channel}
and the measurements are Gaussian.
As a result, the random variables
that determine the mutual information
appear to be normally distributed,
so that we can use the well-known
analytical result that
expresses the mutual information of the
Gaussian random variables
in terms of the determinats of
the suitably defined covariance matrices
(see, e.g., calculations detailed in online book~\cite{SochEtAl2024_StatProofBook}).

Specifically,
transformations that
a coherent state undergoes under the action
of the channel
$\mathcal{N}_{A\to B}$
can be explicitly represented as follows
\begin{align}
  &
  \label{eq:A-to-B-channel}
  \ket{\alpha}\bra{\alpha}\xrightarrow{\mathcal{E}_T}
  \ket{\sqrt{T}\alpha}\bra{\sqrt{T}\alpha}
    \xrightarrow{\Phi_{\xi}}
    N_{\xi}\int\dd^2\beta
    \exp\Bigl\{
    -\frac{2|\beta|^2}{\xi}\Bigr\}
    \notag
  \\
  &
    \times\ket{\sqrt{T}\alpha+\beta}\bra{\sqrt{T}\alpha+\beta},
\end{align}
where $N_\xi=2/(\pi\xi)$.
When Bob performs a Gaussian measurement described by the POVM
$\{\hat{\Pi}_G^{(\gamma)}(x)\}_x$,
where the value of $x$ labels the measurement outcomes
(e.g., for homodyne detection, $x$ is the quadrature variable
given by Eq.~\eqref{eq:x-def}),
the joint probability density can be computed as follows
\begin{subequations}
  \label{eq:P_x-alpha}
\begin{align}
\label{eq:pB(xA)}
  p_{\ind{B}}^{(\gamma)}(x,\alpha) & = p_B^{(\gamma)}(x|\alpha) p_A(\alpha),\quad
                                     \gamma\in\{\ind{H},\ind{DH}\}
  \\
    \label{eq:p_B-cond}
    p_B^{(\gamma)}(x|\alpha)& =\Tr\{\mathcal{E}_{A\to B}(\ket{\alpha}\bra{\alpha})\hat{\Pi}_G^{(\gamma)}(x)\}
    \notag
  \\
  &
    =\Tr\{\mathcal{E}_{T}(\ket{\alpha}\bra{\alpha})\cnj{\Phi}_{\xi}(\hat{\Pi}_G^{(\gamma)}(x))\}
    \notag
  \\
  &
=
    \bra{\sqrt{T}\alpha}\Phi_{\xi}(\hat{\Pi}_G^{(\gamma)}(x))\ket{\sqrt{T}\alpha},
\end{align}
\end{subequations}
where we have used the fact that the additive noise channel $\hat{\Phi}_{\xi}$ is self-dual:
$\cnj{\hat{\Phi}}_{\xi}=\hat{\Phi}_{\xi}$.

For homodyne detection,
we can combine Eq.~\eqref{eq:P_x-alpha}
and the expression for the $Q$ symbol
of POVM given by Eq.~\eqref{eq:Pgood-w-def}
to deduce the joint probability density 
\begin{align}
  &
\label{eq:P-H-xa}
   p_{\ind{B}}^{(\ind{H})}(x,\tilde\alpha_1) 
  \propto
    \exp\left[
    -\frac{(x-2\tilde\alpha_1)^2}{2(\sigma_x+\xi)}-\frac{\tilde\alpha_1^2}{2TV_{\ind{A}}}
  \right]
  \notag
  \\
  &
  =\exp\Bigl\{
  -\frac{1}{2}
    \begin{pmatrix}
        \tilde\alpha_1&x
    \end{pmatrix} (\Sigma_{\ind{H}})^{-1}
    \begin{pmatrix}
        \tilde\alpha_1\\x
    \end{pmatrix}
  \Bigr\},
\end{align}
where
\begin{align}
  \label{eq:talpha}
  \tilde{\alpha}_1=\sqrt{T}\Re\alpha e^{-i\phi},\quad
  \tilde{\alpha}_2=\sqrt{T}\Im\alpha e^{-i\phi}.
\end{align}
The covariance matrix of the normal distribution~\eqref{eq:P-H-xa}
\begin{align}
    \label{eq:sigma-hom-def}
  &
    \Sigma_H=
    \begin{pmatrix}
        {TV_{\ind{A}}}&2{TV_{\ind{A}}}\\
        2{TV_{\ind{A}}}&4TV_{\ind{A}}+\sigma_x+\xi
    \end{pmatrix}
\end{align}
gives  the mutual information
between Alice and Bob that can be computed
from the formula~\cite{SochEtAl2024_StatProofBook}
\begin{align}
\label{eq:IABPM}
    I_{\ind{AB}}^{(\ind{H})}=\frac{1}{2}\log_2 \frac{\Sigma^{\ind{H}}_{11}\Sigma^{\ind{H}}_{22}}
    {\det\Sigma_{\ind{H}}}
    =\frac{1}{2}\log_2\left[1+\frac{4TV_{\ind{A}}}{\sigma_x+\xi}\right],
\end{align}
where $\Sigma^{\ind{H}}_{ii}$ are the diagonal elements of $\Sigma_{\ind{H}}$
and all logarithms are base 2.

Similarly, for double homodyne detection,
from Eqs.~\eqref{eq:P_x-alpha} and~\eqref{eq:P_G-x1x2})
we have
\begin{align}
  \label{eq:dh-exp}
  &
  p_{\ind{B}}^{(\ind{DH})}(x,\tilde \alpha)
    \propto
\prod_{i=1}^2
    \exp\Biggl[
    -\frac{(x_i-\tilde\alpha_i)^2}{\sigma_i+\xi/2}-\frac{\tilde\alpha_i^2}{2TV_{\ind{A}}}
    \Biggr]
                                                \notag
  \\
  &
    =\prod_{i=1}^2
    \exp\Bigl\{
  -\frac{1}{2}
    \begin{pmatrix}
        \tilde\alpha_i&x_i
    \end{pmatrix} (\Sigma_{\ind{DH}}^{(i)})^{-1}
    \begin{pmatrix}
        \tilde\alpha_i\\x_i
    \end{pmatrix}
  \Bigr\}.
\end{align}
Analogously to Eq.~\eqref{eq:sigma-hom-def},
we obtain the diagonal blocks of the covariance matrix
$\Sigma_{\ind{DH}}$
given by
\begin{align}
  \label{eq:Sigma-DH}
  &
    \Sigma_{\ind{DH}}^{(i)}=
    \begin{pmatrix}
        {TV_{\ind{A}}}&{TV_{\ind{A}}}\\
        {TV_{\ind{A}}}&{TV_{\ind{A}}}+(2 \sigma_i+\xi)/4
    \end{pmatrix}, \quad i\in \{1,2\}.
\end{align}

Since
$p_{\ind{B}}^{(\ind{DH})}(x,\tilde\alpha)
=p_{\ind{B}}^{(\ind{DH})}(x_1,\tilde\alpha_1)p_{\ind{B}}^{(\ind{DH})}(x_2,\tilde\alpha_2)
$,
the mutual information can be evaluated as a sum of two terms,
determined by the above formula~\eqref{eq:IABPM}
with $\Sigma_H$ replaced by $\Sigma_{\ind{DH}}^{(i)}$.
The result reads
\begin{align}
  \label{eq:I_dh}
    I_{\ind{AB}}^{(\ind{DH})}&=\frac{1}{2}\sum_{i=1,2}\log_2 \frac{[\Sigma_{\ind{DH}}^{(i)}]_{11}[\Sigma_{\ind{DH}}^{(i)}]_{22}}
    {\det\Sigma_{\ind{DH}}^{(i)}}\notag\\
    &=\frac{1}{2}\sum_i\log_2\left[1+\frac{4TV_{\ind{A}}}
    {2\sigma_i+\xi}\right].
\end{align}

In conclusion of this appendix, we briefly discuss
equivalence relation between the
prepare-and-measure (PM) and the entanglement-based (EB) schemes.
To this end,
we consider the two mode squeezed vacuum (TMSV) state
given by~\cite{Weed:rmp:2012}
\begin{align}
  \label{eq:psi_AB}
  \ket{\Psi_{AB}}=\frac{1}{\cosh r}\sum_{n=0}^{\infty}(\tanh r)^n\ket{n,n}_{\ind{AB}},
\end{align}
where $r$ is the real-valued squeezing parameter,
with the covariance matrix~\cite{Serafini:bk:2023}
\begin{align}
  \label{eq:cov_TMSV}
  &\quad
    \Sigma_{\ind{TMSV}}
  =
  \begin{pmatrix}
        \cosh(2r)\mathbbm{1}&\sinh(2r)\sigma_z\\
        \sinh(2r)\sigma_z&\cosh(2r)\mathbbm{1}
    \end{pmatrix}.
\end{align}

In
the EB protocol,
Alice is assumed to hold the state
$\hat{\rho}_{\ind{AB}}=\ket{\Psi_{AB}}\bra{\Psi_{AB}}$.
Then Alice measures one mode using a noiseless double homodyne measurement,
which corresponds to the coherent-state POVM (see Eq.~\eqref{eq:POVM-hetero-ideal})
$\{\hat\Pi_{CS}(\beta)=\pi^{-1}\ket{\beta}\bra{\beta}\}_{\beta}$,
and sends the second mode to Bob.
Unconditional state on the  Bob's side
prepared by such a measurement 
is
% The probability that Alice observes
% the double homodyne outcome $\beta$ is given by the Born rule:
% \begin{align}
%   \label{eq:p_A_EB}
%   p_{\ind{A}}^{\ind{EB}}(\beta)=\bra{\Psi_{\ind{AB}}} \hat{\Pi}_{CS}(\beta)\otimes\hat{I}_B \ket{\Psi_{\ind{AB}}}
% \end{align}

\begin{align}
  \label{eq:rho_B_EB-1}
  \hat{\rho}_{\ind{B}}^{\ind{EB}}=\int\dd^2\beta \Tr_A \bigl\{(\hat{\Pi}_{CS}(\beta)\otimes\hat{I}_B) \hat{\rho}_{AB}\bigr\}.
\end{align}
By using the relation
\begin{align}
  \label{eq:beta-psi_AB}
  \cosh r \avr{\beta|\Psi_{AB}}=
  \exp\Bigl\{-\frac{|\beta|^2}{2\cosh^2r}\Bigr\} \ket{\tilde{\beta}},
\end{align}
where $\tilde\beta=(\tanh r)\cnj{\beta}$,
the above density matrix~\eqref{eq:rho_B_EB-1}
can be recast into the form
\begin{align}
  \label{eq:rho_B_EB-2}
  \hat{\rho}_{\ind{B}}^{\ind{EB}}=
\frac{1}{\pi \sinh^2 r}
  \int\dd^2\tilde{\beta}
  \exp\Bigl\{-\frac{|\tilde{\beta}|^2}{\sinh^2 r}\Bigr\}
  \ket{\tilde{\beta}}\bra{\tilde{\beta}}.
\end{align}
When
\begin{align}
  \label{eq:var-equiv}
  \sinh^2 r = 2 V_A
\end{align}
and $\tilde\beta$ is changed to $\alpha$,
the expressions for
the input density matrix
in the PM picture~\eqref{eq:ensemble}
and
for the density matrix after Alice's measurement
in the EB picture~\eqref{eq:rho_B_EB-2}
become identical.
Since relation
$\cosh(2r)=1+4V_A\equiv V$ immediately follows from Eq.~\eqref{eq:var-equiv},
the covariance matrix~\eqref{eq:cov_TMSV}
equals
the matrix $\Sigma_{\ind{TMSV}}$
that enters Eq.~\eqref{eq:TMSVS-to-AB}.
This results justify using
the covariance matrix of TMSV
for calculation of the Holevo information in the PM protocol.

%\bibliography{quant,refs,homodyne,math}


\end{document}